\documentclass[journal]{IEEEtran}
\usepackage{amsbsy,amsmath,amssymb,epsfig,bbm,mathrsfs,fancyhdr,fancyvrb,subfigure,url,cite,multirow,array}
\usepackage{graphicx,bm,makecell}
\usepackage{epstopdf}
\usepackage{setspace}
\usepackage[ruled,linesnumbered]{algorithm2e}
\allowdisplaybreaks

\usepackage{textcomp,booktabs}
\usepackage[usenames,dvipsnames]{color}
\usepackage{colortbl}
\definecolor{mygray}{gray}{.93}
\definecolor{mycyan}{cmyk}{.09,0,0,0}
\definecolor{mypink}{rgb}{.96,.91,.95}
\definecolor{mygray1}{gray}{.85}
\usepackage[table,xcdraw]{xcolor}

\newcommand{\beq}{\begin{equation}}
\newcommand{\eeq}{\end{equation}}
\newcommand{\beqn}{\begin{eqnarray}}
\newcommand{\eeqn}{\end{eqnarray}}

\usepackage{amsmath}
\newtheorem{thm}{\protect\theoremname}

\newtheorem{myDef}{\protect\Definition}
\providecommand{\theoremname}{\textbf{Theorem}}
\providecommand{\propositionname}{\textbf{Proposition}}
\providecommand{\remarkname}{\textbf{Remark}}
\providecommand{\lemmaname}{\textbf{Lemma}}
\providecommand{\corollaryname}{\textbf{Corollary}}
\providecommand{\Definition}{\textbf{Definition}}

\begin{document}

\title{Approximated Coded Computing: Towards Fast, Private and Secure Distributed Machine Learning}
\author{Houming Qiu, Kun Zhu,~\IEEEmembership{Member,~IEEE}, Nguyen Cong Luong, and Dusit Niyato,~\IEEEmembership{Fellow,~IEEE}
\thanks{This paper was presented in part at the CollaborateCom 2023\cite{qiu2023secure}.}
\thanks{H. M. Qiu and K. Zhu are with the College of Computer Science and Technology, Nanjing University of Aeronautics and Astronautics, Nanjing 210016, China (email: \{hmqiu56, zhukun\}@nuaa.edu.cn).}
\thanks{N. C. Luong is with the Faculty of Computer Science, Phenikaa University, Hanoi 12116, Vietnam (email: luong.nguyencong@phenikaa-uni.edu.vn).}
\thanks{D. Niyato is with School of Computer Science and Engineering, Nanyang Technological University, Singapore 639798 (email: dniyato@ntu.edu.sg).}
} \maketitle

\begin{abstract}
In a large-scale distributed machine learning system, coded computing has attracted wide-spread attention since it can effectively alleviate the impact of stragglers. However, several
emerging problems greatly limit the performance of coded distributed systems. Firstly, an existence of colluding workers who collude results with each other leads to serious privacy leakage
issues. Secondly, there are few existing works considering security issues in data transmission of distributed computing systems/or coded distributed machine learning systems. Thirdly, the
number of required results for which need to wait increases with the degree of decoding functions. In this paper, we design a secure and private approximated coded distributed computing (SPACDC) scheme that deals with the above-mentioned problems simultaneously. Our SPACDC scheme guarantees data
security during the transmission process using a new encryption algorithm based on elliptic curve cryptography. Especially, the SPACDC scheme does not impose strict
constraints on the minimum number of results required to be waited for. An extensive performance analysis is conducted to demonstrate the effectiveness of our SPACDC scheme.
Furthermore, we present a secure and private distributed learning algorithm based on the SPACDC scheme, which can provide information-theoretic privacy protection for training data. Our
experiments show that the SPACDC-based deep learning algorithm achieves a significant speedup over the baseline approaches.
\end{abstract}


\section{Introduction}
\IEEEPARstart{M}{achine} learning (ML) has made significant advancements in recent years, leading to an extraordinary era of intelligence. ML algorithms, especially deep learning (DL)
algorithms, are being increasingly applied to improve the quality of our daily lives in several areas, including smart homes~\cite{dong2022multimodal}, face recognition~\cite{qiu2021end2end},
recommender systems~\cite{yu2024self}, and more. However, the exponential increase in dataset size and neural model complexity results in an unbearable amount of time required for the
training process. For instance, it takes around two days to train a widely used ResNet-$50$~\cite{yu2020resnet} model on the ImageNet dataset for $90$ epochs using the most recent NVIDIA Tesla V100 GPU~\cite{yamamoto2020parallel,zhang2022picasso}. Certainly, a single computing machine is unable to catch up to the training demands of extremely-high computing and storage resources while achieving the anticipated training time for emerging large-scale model training tasks, especially like chatGPT.

Distributed machine learning (DML) systems emerge as a class of promising computing paradigms in response to the above problems~\cite{hu2021distributed}. Generally, DML systems are implemented
in the form of a parameter server framework to train a large ML model using several collaborative workers. The parameter server framework is composed of a server along with multiple clients, where the
server plays the role of a master for aggregating the weights from clients, and clients as workers complete training tasks in parallel~\cite{bao2022deep}. Therefore, the total model training time is dramatically reduced, and the heavy burden of computing and storage resources for a single computing machine are significantly alleviated. However, network congestion, asymmetric computational load, and resource heterogeneity inevitably lead to the existence of stragglers in the DML systems~\cite{amiri2019computation,buyukates2022gradient,wang2023fsp}. Stragglers refer to some faulty clients, or some clients unpredictably return computed results extremely slower than other clients.

Coded distributed computing (CDC) is regarded as an efficient technique for mitigating the impact of stragglers, which has attracted considerable
attention~\cite{yazdanialahabadi2022distributed,lu2022decode,ozfatura2022coded,qiu2023secure}. It injects coding theory into DML frameworks to enhance resilience against stragglers. To be more specific,
using the CDC technique, the server only needs to utilize the weights returned by any subset of all clients to recover the final aggregated weight. Recently, numerous studies have proved that
the CDC technique is excellent for accelerating the model training process in DML systems~\cite{li2023data,so2021codedprivateml,Jahani2023Berrut}. Due to the fact that the datasets required to
train an ML model contain sensitive information about users or organizations, such as personally identifiable information, social relationships, habits, and preferences, utilizing the raw datasets for model training can easily lead to information leakage~\cite{tang2022adaptive}. Especially the colluding clients (workers), i.e., clients that coordinate with each other to gain more sensitive information, further weaken the privacy protection of the datasets. Therefore, it is crucial to train a large-scale model based on the DML framework employing CDC technique while guaranteeing data privacy and security.

In~\cite{so2021codedprivateml,Jahani2023Berrut,tang2022adaptive}, the authors designed CDC schemes for DML to speed up the model training process while considering privacy-preserving of the
datasets. These proposed CDC schemes provide privacy protection for the raw datasets by adding a set of random matrices. However, the addition of random matrices greatly raises  the recovery threshold, which is the minimum number of clients needed to return weights back to the server~\cite{yu2020entangled}. This inevitably prolongs the training time of a large-scale ML model. Furthermore, the recovery threshold is strictly constrained. It indicates that the server cannot recover the final weight until it has received a sufficient number of returned weights from the clients. Hence, the CDC schemes need to be designed to guarantee the privacy and security of training datasets while relaxing the strict constraints on the recovery threshold.

On the other hand, network communication in distributed computing systems is vulnerable to various potential threats and weaknesses~\cite{chatterjee2022physically}. As a result, data is
susceptible to being eavesdropped during the transmission process, causing privacy and security issues. Unfortunately, most of the CDC-related
studies~\cite{lee2018speeding,Yu2017Polynomial,dutta2020optimal,yu2020straggler,jia2021cross,yu2019lagrange,tauz2022variable,wang2021batch,hasirciouglu2021bivariate,Das2022Coded,Das2023Distributed,zhu2023information,yang2019secure,aliasgari2020private,li2022private,zhu2021secure,zhu2022symmetric,zhu2022generalized,yao2023jamming}
mainly focus on the data privacy and security issues caused by the clients while disregarding the data transmission process. In addition, elliptic curve cryptography (ECC) is characterized by
high security with smaller key sizes, leading to more efficient computation and less bandwidth for communication networks~\cite{chaudhry2022sg}. This motivates us to study a CDC scheme to
guarantee security and privacy of the training datasets while addressing the security problems of the data transmission process.

As mentioned above, existing studies have several limitations. Firstly, CDC-based schemes extend to data privacy protection, leading to a significant increase in the recovery threshold.
Secondly, they are restricted to a specific type of task, e.g., matrix multiplication. Thirdly, existing decoding methods, such as polynomial interpolation algorithm and inverse of the coefficient matrix, strictly limit the recovery threshold. Fourthly, the security problem of data during the transmission process is not considered. To tackle the aforementioned problems, we aim to design an efficient CDC scheme to shorten the execution time of ML model training while preserving the privacy and security of the datasets. The main contributions of this paper are listed as follows:
\begin{itemize}
\item [$\bullet$]
We address the limitation on the recovery threshold and propose a secure and private approximated coded distributed computing (SPACDC) scheme. In particular, our proposed scheme relaxes the strict limitations on the minimum number of results that clients must return while achieving resiliency against stragglers, and enables information-theoretic privacy (ITP) protection against colluding workers.
\item [$\bullet$]
We fill the gap between CDC systems and network transmission and introduce a MEA-ECC algorithm leveraging ECC for a distributed system, designed to secure data during the transmission process.
\item [$\bullet$]
We design a private and secure distributed learning algorithm based on the proposed SPACDC scheme, named the SPACDC-DL algorithm. Compared to the baseline algorithms, our SPACDC-DL algorithm
can recover the aggregated weight parameter without waiting for all results from clients while offering privacy and security protection for the datasets. The experiments show the
excellent performance of the SPACDC-DL algorithm.
\item [$\bullet$]
Finally, the SPACDC scheme is theoretically proved to ensure the ITP of the input data. Moreover, the superiority of our SPACDC scheme is demonstrated by extensive performance evaluation.
\end{itemize}

The rest of this paper is organized as follows. We provide the related works in Section~\uppercase\expandafter{\romannumeral2}. Then, we present the system model and problem formulation in Section~\uppercase\expandafter{\romannumeral3}. In Section~\uppercase\expandafter{\romannumeral4}, we introduce a new matrix encryption algorithm. After that, we propose the SPACDC scheme in Section \uppercase\expandafter{\romannumeral5}. Section \uppercase\expandafter{\romannumeral6} presents the SPACDC-DL algorithm. In Section \uppercase\expandafter{\romannumeral7}, we discuss  the convergence and provide experiments for our proposed SPACDC-DL. In Section \uppercase\expandafter{\romannumeral8}, we provide comprehensive complexity analysis for the SPACDC scheme. Finally, we conclude the paper in Section \uppercase\expandafter{\romannumeral9}.

The abbreviations appeared in the paper are listed in Table~\ref{table1}.

\begin{table}
\center
\caption{Major abbreviations}
\label{table1}
\footnotesize
\begin{tabular}{|p{40pt}|p{170pt}|}
\hline
\makecell*[l]{Abbreviations}   & \makecell*[l]{Description} \\
\hline
$\mathbf{a}^l$ & Output vector of the $l$th layer\\
\hline
$\mathbf{b}$  & Bias vector\\
\hline
$\mathbf{C}_i$& Ciphertext of $\mathbf{X}_i$\\
\hline
$\tilde{\mathbf{C}}_i$& Ciphertext of $\tilde{\mathbf{Y}}_i$\\
\hline
$\mathcal{D}$ & Training dataset\\
\hline
$G$ & Generator point\\
\hline
$K$    & Number of submatrices\\
\hline
$\mathcal{K}$ & Set of indexes of returned workers\\
\hline
$L$    & Number of Layer for DNN\\
\hline
$\mathbf{M}_i$ & A confidential matrix\\
\hline
$N$   & Number of workers \\
\hline
$\mathcal{N}$ & Set of indexes of the $N$ workers\\
\hline
$P$ & A point $P(x_1,y_i)$\\
\hline
$\mathcal{P}$   &Set of colluding workers\\
\hline
$pk_M$   &The master's public key\\
\hline
$pk_{W_i}$   &$W_i$'s public key\\
\hline
$Q$ & A point $Q(x_2,y_2)$\\
\hline
$\mathcal{S}$   &Set of stragglers\\
\hline
$S$    & Number of stragglers\\
\hline
$sk_M$  &The master's private key\\
\hline
$sk_{W_i}$  &$W_i$'s Private key\\
\hline
$s_{K_i}$    &Share key\\
\hline
$T$   & Number of colluding workers\\
\hline
$V$ & A point $V(x_3,y_3)$\\
\hline
$W_i$ & Worker with index $i$\\
\hline
$\mathbf{X}$ & Input dataset\\
\hline
$\tilde{\mathbf{X}}_i$ & Encoded submatrix\\
\hline
$\mathbf{Y}$   & Final result\\
\hline
$\tilde{\mathbf{Y}}_i$   & $W_i$'s sub-result\\
\hline
$\mathbf{Z}_i$   & Random matrix\\
\hline
$\mathbf{\Theta}$  & Weight matrix \\
\hline
\end{tabular}
\end{table}

\section{Related Work}
\subsection{Coded Distributed Computing}
In master-worker computing systems, stragglers refer to workers who are extremely slow in completing a computational task~\cite{Jahani2023Berrut}. The existence of stragglers greatly
extends the total waiting time of the task. Recently, CDC has gained extensive attention due to its ability to effectively alleviate the straggler effects.
In~\cite{lee2018speeding}, the authors dealt with the problem of straggler effects for distributed matrix multiplication (DMM) using the maximum distance separable (MDS) codes. This opens a
promising research direction toward mitigating straggler effects for large-scale DMM tasks. In~\cite{Yu2017Polynomial}, the authors proposed polynomial codes for the multiplication
of two high-dimensional matrices $\mathbf{A}$ and $\mathbf{B}$. More precisely, the master partitions $\mathbf{A}$ and $\mathbf{B}$ by row and column, respectively. Then, two encoding functions are
carefully designed, utilizing the partitions of $\mathbf{A}$ and $\mathbf{B}$. On the contrary, the authors in~\cite{dutta2020optimal} proposed MatDot codes, which partition $\mathbf{A}$ and
$\mathbf{B}$ by column and row, respectively. The recovery threshold is lower for the MatDot codes compared to the polynomial codes. However, the communication and computation costs of the
MatDot codes are higher than those of the polynomial codes. To balance the recovery thresholds and computation/communication costs, the authors in~\cite{dutta2020optimal} also proposed a class
of unified coding schemes, i.e., PolyDot codes, which offer a tradeoff between the recovery thresholds and computation/communication loads. In fact, the PolyDot codes are the combination of
both the polynomial codes and the MatDot codes.

In contrast to~\cite{Yu2017Polynomial} and~\cite{dutta2020optimal}, the authors in~\cite{yu2020straggler} proposed an entangled polynomial (EP) code, which further reduces the recovery threshold. In the EP code, two high-dimensional matrices $\mathbf{A}$ and $\mathbf{B}$ are both divided by column and row. In~\cite{jia2021cross}, the authors proposed cross-subspace alignment
(CSA) codes for batch processing of matrices in distributed systems. The CSA codes outperform the Lagrange coded computing (LCC)~\cite{yu2019lagrange} scheme under limited downloads,
where the LCC scheme encodes data by leveraging the famous Lagrange polynomial. Inspired by the CSA codes, the authors in~\cite{tauz2022variable} proposed flexible CSA codes, i.e., FSCA codes,
which divide the target computational task into multiple groups and then carefully design encoding functions based on rational functions to reduce the interference of unexpected computations.
The FCSA codes achieve flexible utilization of computational redundancy while offering strong robustness against stragglers. In~\cite{wang2021batch,hasirciouglu2021bivariate,Das2022Coded,Das2023Distributed}, the authors designed coding schemes to alleviate the straggler effects while reducing the computation time by leveraging the work done by stragglers. However, the recovery thresholds of these proposed coding schemes are vulnerable to increase with the degree of the encoding functions. In addition, the above coding schemes cannot protect ITP of the input data.

\subsection{Coded Distributed Computing with Privacy Protection}
To provide privacy protection for the confidential data, the authors in~\cite{zhu2023information} improved the MDS-coded scheme~\cite{lee2018speeding} for DMM tasks. The master prevents
colluding workers from obtaining information from input matrices by adding a set of randomly generated matrices. Similarly, the authors in~\cite{yang2019secure} extended the polynomial
codes~\cite{Yu2017Polynomial} to secure polynomial codes. However, this coding scheme cannot tolerate the existence of colluding workers. To overcome this problem, the authors in~\cite{aliasgari2020private} proposed secure generalized PolyDot (SGPD) codes, which improve the PolyDot codes~\cite{dutta2020optimal} in ensuring the data privacy against $P_{C}$ colluding workers. In~\cite{li2022private}, the authors proposed two private and secure DMM coding schemes, which obtain a smaller download cost than that of the SGPD codes. In~\cite{qiu2024coded}, the authors dealt with the problem of privacy protection and designed a resilient, secure, and private coded scheme for the edge-enable Metaverse. We recognize that these coding schemes~\cite{zhu2023information,yang2019secure,aliasgari2020private,li2022private,qiu2024coded} keep the dataset private by injecting a set of random matrices to mask the raw data. However, the recovery thresholds of these coding schemes are too large, which indicates that the master is required to collect more computed results from workers.

In~\cite{yu2020entangled,zhu2021secure,zhu2022symmetric,zhu2022generalized}, the authors improved the LCC~\cite{yu2019lagrange} scheme with respect of evaluating an arbitrary multivariate
polynomial in distributed computing while providing resiliency against stragglers and privacy protection for the dataset. Although the aforementioned coding schemes enable data privacy against a certain number of colluding workers, they cannot recover the final result unless the number of returned computed results meets the recovery threshold. Furthermore, most previous studies are largely concerned with data security at the worker level, while disregarding the security concerns associated with transmitting data. Meanwhile, data is vulnerable to eavesdropping during its transmission in real-world circumstances~\cite{yao2023jamming}.

To deal with the above-mentioned issues, we focus on designing a novel coding scheme, i.e., the SPACDC scheme that (\textit{i}) guarantees ITP against a variable number of colluding workers, (\textit{ii}) provides robustness against stragglers, and (\textit{iii}) prevents data from eavesdropping during the transmission process. Moreover, we integrate the SPACDC scheme with distributed deep learning to accelerate the training process while guaranteeing the privacy and security of the training datasets.

\section{System Model and Problem Formulation}
In this section, we introduce a CDC system model that incorporates encrypted communication. In addition, several important definitions relevant to this work are given. After that, we formulate the problem of ensuring security and privacy in large-scale computing within a CDC system.

\subsection{System Model}
\begin{figure}[!t]
\centering
\includegraphics[width=3.2in]{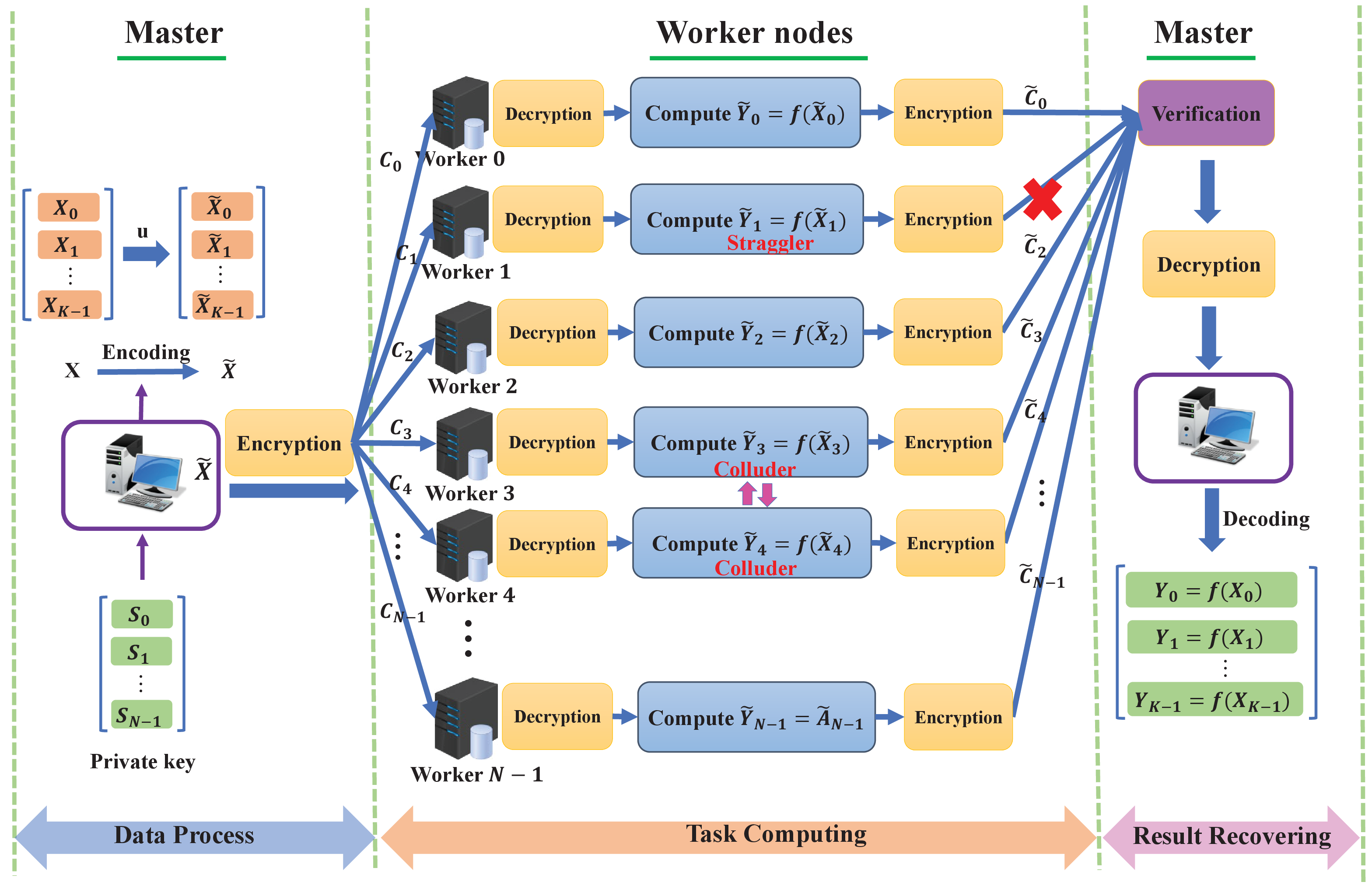}
\caption{An illustration of a CDC system based on the proposed matrix encryption algorithm. Therein, the master aims to approximately compute a polynomial over a dataset
$\mathbf{X}=[\mathbf{X}^T_0,\mathbf{X}^T_1,\ldots,\mathbf{X}^T_{K-1}]^T$. }
\label{fig:system-model}
\end{figure}

We consider a CDC system with a master and a set $\mathcal{N}$ of $N$ workers. Worker $i \in \mathcal{N}$ is denoted by $W_i$. In the system, the number of straggling and colluding workers is $S$ and $T$, respectively. The master aims to achieve an approximation of evaluating a function $f:\mathbb{V}\rightarrow\mathbb{U}$ over a high-dimensional dataset
$\mathbf{X}$, i.e., $\mathbf{Y}\approx f(\mathbf{X})$, where $\mathbf{X}$ has a dimension of $m\times d$, $d>0$ and $m>0$ are two integers, $\mathbb{V}$ and $\mathbb{U}$ are the two real matrix spaces. Specifically, this system model is potentially vulnerable to eavesdroppers during the network transmission process. As illustrated in Fig.~\ref{fig:system-model}, we detail the entire encoding and decoding procedure as follows:

\begin{enumerate}
\item
\emph{\textbf{Data Process:}}
The dataset $\mathbf{X}$ is divided into multiple submatrices, which are then encoded into $N$ encoded matrices. The encoded matrix corresponding to worker $i$ is denoted by $\tilde{\mathbf{X}}_i$, i.e.,
\begin{equation*}
\begin{split}
\tilde{\mathbf{X}}_i=\mathrm{g}_i(\mathbf{X})~\text{for}~i\in\mathcal{N}\triangleq\{0,1,\ldots,N-1\},
\end{split}
\end{equation*}
where $\tilde{\mathbf{X}}_i$ has a dimension of $\frac{m}{K}\times d$, integer $K>0$, and $\mathrm{g}_i$ is the encoding function defined by: $\mathrm{g}_i:\mathbb{F}^{m\times d}\rightarrow\mathbb{F}^{\frac{m}{K}\times d}$. $\mathbb{F}$ is a sufficiently large field. To enhance data security, $\tilde{\mathbf{X}}_i$ is encrypted into ciphertext $\mathbf{C}_i$ by utilizing the MEA-ECC algorithm by the master. Then, ciphertext $\mathbf{C}_i$ for $i\in\mathcal{N}$ is assigned to worker $W_i$ by the master.
\item
\emph{\textbf{Task Computing:}}
Worker $W_i$ uses its private key to decrypt the encrypted data $\mathbf{C}_i$ that it received, thus retrieving the original data $\tilde{\mathbf{X}}_i$. Then, worker $W_i$ executes the
task $\tilde{\mathbf{Y}}_i=f(\tilde{\mathbf{X}}_i)$. Once finishing the assigned task, worker $W_i$ follows step 3 of the MEA-ECC algorithm to encrypt the calculated result $\tilde{\mathbf{Y}}_i$ into $\tilde{\mathbf{C}}_i$, which is then sent back to the master. It should be noted that some workers may not be able to complete the assigned tasks or take longer than others to send $\tilde{\mathbf{Y}}_i$ back to the master.
\item
\emph{\textbf{Result Recovering:}}
The master receives $\{\tilde{\textbf{C}}_i\}_{i\in\bm{\mathcal{K}}}$ from worker $W_i$ and records its index with $\mathcal{K}\in\mathcal{N}$. Then, the master uses the MEA-ECC algorithm to decrypt $\tilde{\textbf{C}}_i$ and recovers the original calculation results $\tilde{\textbf{Y}}_i$ for $i\in\bm{\mathcal{K}}$. Following that, the final result $\mathbf{Y}$ can be obtained by utilizing the following function:
\begin{equation}
\begin{split}
\mathbf{Y}=\hslash_{\mathcal{K}}\bigg(\{\tilde{\mathbf{Y}}_i\}_{i\in\mathcal{K}}\bigg).
\end{split}
\end{equation}
\end{enumerate}

\subsection{Related Definitions}
\begin{myDef}
In a CDC system, the privacy constraint of encoded data $\tilde{\mathbf{X}}_{\mathcal{P}}$ that is sent to worker $\{W_i\}_{i\in\mathcal{P}}$ is given by:
\begin{equation}
\begin{split}
\mathbf{I}(\tilde{\mathbf{X}}_\mathcal{P};\mathbf{X})&=0,
\end{split}
\label{eqPri}
\end{equation}
where $\mathcal{P}$ represents the set of indices for the $T$ colluding workers and $\mathbf{I}(\bullet;\bullet)$ represents the mutual information.

\label{DefPrivacy}
\end{myDef}

\begin{myDef}
Given a finite field $\mathbb{F}_{q}$, the elliptic curve equation is given by:
\begin{equation}
\label{ECequation}
y^2=x^3+ax+b,
\end{equation}
where the coefficients $a$, $b\in\mathbb{F}_q$ satisfy the following condition:
\begin{equation}
4a^3+27b^2\neq 0.
\end{equation}
\end{myDef}

\begin{myDef}
Given a function $f$ and $n$ distinct points $(x_0,f_0),\ldots,(x_{n-1},f_{n-1})$, where $a<x_i<b$ for $i\in\{0,1,\ldots,n-1\}$. The Berrut's rational interpolant of $f$ is given by:
\begin{equation}
r(x)\triangleq\sum_{i=0}^{n-1}f_il_i(x),
\end{equation}
where $l_i(x)$ is the basic function of the form
\begin{equation}
l_i(x)\triangleq\frac{\frac{(-1)^i}{x-x_i}}{\sum_{i=0}^{n-1}\frac{(-1)^i}{x-x_i}}.
\end{equation}
\end{myDef}

\subsection{Problem Formulation}
This paper explores an approximation of evaluating a function over a high-dimensional dataset in a distributed computing scenario with a master and $N$ workers based on coding theory. In the
system, the number of straggling and colluding workers is $S$ and $T$, respectively. Straggling and colluding workers inherently lead to long waiting time and privacy leakage, respectively. In
addition, the emerging communication eavesdropping issues lead to unreliable communication, compromising data security.

To tackle the above-mentioned problems, we focus on designing a new CDC scheme using ECC to achieve lower decoding complexity while considering the presence of stragglers, colluding workers, and eavesdroppers. To be more specific, this scheme not only prevents data from being eavesdropped during transmission but also lowers the recovery threshold and provides robust ITP protection. Furthermore, the proposed scheme extends to enable fast, secure, and private model training of deep neural networks (DNNs) in a distributed computing framework.

\section{Matrix Encryption Algorithm Based on Elliptic Curve Cryptography}
In this section, we introduce the MEA-ECC, a matrix encryption algorithm designed for a distributed system using ECC. Firstly, we discuss ECC, a cost-efficient and highly secure encryption method.

\subsection{Overview of ECC}
ECC is a class of elliptic curve theory-based asymmetric public key encryption methods. The key idea of ECC is to utilize point operations on an elliptic curve to provide strong security encryption. The elliptic curve over a finite field $\mathbb{F}_q$ has the form of
\begin{equation}
y^2=\{x^3+ax+b\}~\text{mod}~\{q\},
\end{equation}
which is called Weierstrass equation, the discriminant is defined as follows:
\begin{equation}
\{4a^3+27b^2\}~\text{mod}~\{q\}\neq 0.
\end{equation}

ECC provides point addition and point multiplication, two basic mathematical operations. The specific execution procedures of these two mathematical operations can be found in our previous work~\cite{qiu2023secure}. 
\begin{enumerate}
\item
\emph{\textbf{Point Addition or Doubling:}}
Given two points, $P(x_1,y_1)$ and $Q(x_2,y_2)$ on the elliptic curve, the addition of $P$ and $Q$ is defined as $P+Q=V(x_3,y_3)$, where
\begin{equation}
x_3=\{\lambda^2-x_1-x_2\}~\text{mod}~\{q\},
\end{equation}
\begin{equation}
y_3=\{\lambda(x_1-x_3)-y_1\}~\text{mod}~\{q\},
\end{equation}
and
\begin{equation}
\lambda=\begin{cases}
\{\frac{y_2-y_1}{x_2-x_1}\}~\text{mod}~\{q\},&~\text{if}~P\neq Q;\\
\{\frac{3x_1^2+a}{2y_1}\}~\text{mod}~\{q\},&~\text{if}~P=Q.
\end{cases}
\end{equation}
\item
\emph{\textbf{Point Multiplication:}}
Given an integer $i>0$ and any point $Q$ on the elliptic curve, the point multiplication $i\cdot Q$ is defined as
\begin{equation}
i\cdot Q=\underbrace{Q+Q+\cdots+Q}_{i~\text{times}},
\end{equation}
\end{enumerate}

\subsection{MEA-ECC Algorithm}
In this subsection, we present the MEA-ECC algorithm. By using the MEA-ECC algorithm, a confidential matrix $\mathbf{M}_i$ of dimension $m\times d$ can be sent from the master to worker $W_i$ in a secure manner. The MEA-ECC's specific encryption procedures are as follows:
\begin{enumerate}
\item
\emph{\textbf{Key generation:}}
The master selects a random integer $sk_M<q$ as its private key, and then executes $pk_M=sk_M\cdot G$ to obtain the public key, where $G$ represents the generator point. Similarly, worker $W_i$ selects $sk_{W_i}<q$ and executes $pk_{W_i}=sk_{W_i}\cdot G$ as its private key and public key, respectively.
\item
\emph{\textbf{Key Exchanging:}}
The master executes $s_{K_i}=sk_M\cdot pk_{W_i}$ to obtain the share key. Similarly, worker $W_i$ executes $s_{K_i}'=sk_{W_i}\cdot pk_M$ to obtain the share key. Certainly, we have $s_{K_i}=sk_M\cdot pk_{W_i}=sk_M(sk_{W_i}\cdot G)=sk_{W_i}(sk_M\cdot G)=sk_{W_i}\cdot pk_M=s_{K_i}'$.
\item
\emph{\textbf{Encryption:}}
Worker $W_i$ receives confidential matrix $\mathbf{M}_i$ from the master. We define $\mathbf{C}_{i}=\{k\cdot G,~\mathbf{M}_i+\Psi(k\cdot pk_{W_i})\mathbf{I}_{m,d}\}$ as a ciphertext point, where $\Psi(x,y)=x$. $\mathbf{I}_{m,d}\in\mathbb{F}^{m\times d}$ is a all-ones matrix, and integer $k$ satisfies $1<k<q$.
\item
\emph{\textbf{Decryption:}}
Worker $W_i$ executes $\mathbf{M}_i+\Psi(k\cdot pk_{W_i})\mathbf{I}_{m,d}-\Psi[sk_{W_i}(k\cdot G)]\mathbf{I}_{m,d}=\mathbf{M}_i+\Psi[k(sk_{W_i}\cdot G)-sk_{W_i}(k\cdot G)]\mathbf{I}_{m,d}=\mathbf{M}_i$ to obtain original data $\mathbf{M}_i$.
\end{enumerate}

\section{The Proposed Coding Scheme}
In this section, we propose an approximated CDC scheme using ECC, i.e., the SPACDC scheme. An example is given to show the key idea of our SPACDC scheme before a general description is introduced.

\subsection{Illustrating Example}
Consider an approximated distributed computing task of the function $f(\mathbf{X}_i)=\mathbf{X}_i\mathbf{X}^T_i$ using $N=8$ workers. Given system parameters: $S=T=1$ and $K=2$.
The master equally splits the high-dimensional matrix $\mathbf{X}$ into $K=2$ block matrices as
\begin{equation}
\mathbf{X}=
\begin{bmatrix}
\mathbf{X}_{0}' \\
\mathbf{X}_{1}'
\end{bmatrix},
\end{equation}
where $\mathbf{X}_{0}'\in\mathbb{F}^{\frac{m}{2}\times d}$ and $\mathbf{X}_{1}'\in\mathbb{F}^{\frac{m}{2}\times d}$. Thus, the master's task is transformed into an approximation of computing  $f(\mathbf{X}_i')$ for $i=0,1$.

Firstly, the submatrices $\mathbf{X}_{0}'$ and $\mathbf{X}_{1}'$ of the input data $\mathbf{X}$ are encoded by using the designed encoding function as follows:
\begin{equation}
\begin{split}
\varrho(x)=\frac{1}{(x-1)\ell(x)}\mathbf{X}_0'-\frac{1}{(x-2)\ell(x)}\mathbf{X}_1'+\frac{1}{(x-3)\ell(x)}\mathbf{Z}_0',
\end{split}
\label{encodingFunc}
\end{equation}
where $\ell(x)=1/(x-1)-1/(x-2)+1/(x-3)$. The random matrix $\mathbf{Z}_0'\in\mathbb{F}^{\frac{m}{2}\times d}$ is generated with independent and identically distributed (i.i.d.) uniform random elements from $\mathbb{F}$. It is worth noting that $\varrho(1)=\mathbf{X}_0'$, $\varrho(2)=\mathbf{X}_1'$ and $\varrho(3)=\mathbf{Z}_0'$. In addition, $8$ distinct values $\{\alpha_i'\}_{i=0}^{7}$ over field $\mathbb{F}$ need to be selected while satisfying $\{\alpha_i'\}_{i=0}^{7}\cap\{1,2,3\}=\varnothing$.
Hence, the master obtains the encoded data $\{\tilde{\mathbf{X}}_i'=\varrho(\alpha_i')\}_{i=0}^{7}$ by utilizing Eq~\eqref{encodingFunc}.

To prevent data from being eavesdropped during transmission, the encoded data $\{\tilde{\mathbf{X}}_i'=\varrho(\alpha_i)\}_{i=0}^{7}$ can be encrypted into ciphertext $\{\mathbf{C}_i'\}_{i=0}^{7}$ by
the master using the MEA-ECC. Following that, the master distributes the $\mathbf{C}_i'$ to worker $W_i$, for $i\in\{0,1,2,\ldots,7\}$. Once the $\mathbf{C}_i'$ is received by
worker $W_i$, $\mathbf{C}_i'$ is decrypted using its private key to obtain the original coded data $\tilde{\mathbf{X}}_i'$.

After that, worker $W_i$ executes the assigned task $\tilde{\mathbf{Y}}_i'=f(\tilde{\mathbf{X}}_i')$. Before sending the computation result back to the master, the worker uses the MEA-ECC algorithm
to encrypt the result $\tilde{\mathbf{Y}}_i'$ into ciphertext $\tilde{\mathbf{C}}_i'$. This encryption phase is crucial for guaranteeing the security of computation results during transmission.
After the master receives the $\tilde{\mathbf{C}}_i'$ from worker $W_i$, it decrypts this data using its private key to recover the original result $\tilde{\mathbf{Y}}_i'$. Therefore, inspired by Berrut's rational interpolant~\cite{Jahani2023Berrut}, we design the following decoding function $\hbar(x)$ based on received points $(\alpha_i',\tilde{\mathbf{Y}}_i')$, is given by
\begin{equation}
\begin{split}
\hbar(x)=\sum_{i\in\mathcal{F}'}\frac{\frac{(-1)^i}{x-\alpha_i'}}{\sum_{j\in\mathcal{F}'}\frac{(-1)^j}{x-\alpha_j'}}\tilde{\mathbf{Y}}_i',
\end{split}
\label{eqh}
\end{equation}
where $i\in\mathcal{F}'$ and $\mathcal{F}'$ represents the set of indexes for the workers who complete their assigned tasks.

By substituting $x=1$ and $x=2$ into the above equation, we can obtain an approximation of $f(\mathbf{X}_0')$ and $f(\mathbf{X}_1')$, respectively. Consequently, the master accomplishes the computational task.

\subsection{General SPACDC Scheme Design}
In this subsection, we introduce our proposed SPACDC scheme in a general setting. The SPACDC scheme mainly consists of three phases, as described below:
\subsubsection{\textbf{Data Process}}
The master equally partitions the high-dimensional matrix $\mathbf{X}\in\mathbb{F}^{m\times d}$ into $K$ blocks of submatrices by row, i.e.,
\begin{equation}
\label{eqDivMtx}
\mathbf{X}=
\begin{bmatrix}
\mathbf{X}_0 \\
\mathbf{X}_1 \\
\vdots\\
\mathbf{X}_{K-1}
\end{bmatrix},
\end{equation}
where $\mathbf{X}_i\in\mathbb{F}^{\frac{m}{K}\times d}$ for $i\in\{0,1,\ldots,K-1\}$ and integer $K>0$. If $m$ is not divisible by $K$, the final block may be zero-padded. Then, the master's task is transformed into computing an approximation of $\mathbf{Y}_i\approx f(\mathbf{X}_i)$ for $i\in\mathcal{K}$.

On the other hand, we select any $K+T$ distinct values $\beta_1,\beta_2,\ldots,\beta_{K+T-1}$ from field $\mathbb{F}$. After that, the submatrices blocks $\{\mathbf{X}_0,\mathbf{X}_1,\ldots,\mathbf{X}_{K-1}\}$ are encoded by the master utilizing the designed function as follows:
\begin{equation}
\begin{split}
\label{eqEndFunc}
u(z)=\sum_{i=0}^{K-1}\frac{(-1)^i}{(z-\beta_i)\Gamma(z)}\mathbf{X}_i+\sum_{i=K}^{K+T-1}\frac{(-1)^i}{(z-\beta_i)\Gamma(z)}\mathbf{Z}_i,
\end{split}
\end{equation}
where $\Gamma(z)=\sum_{j=0}^{K+T-1}\frac{(-1)^j}{z-\beta_j}$. Random matrices $\{\mathbf{Z}_{i}\}_{i=K}^{K+T-1}\in\mathbb{F}^{\frac{m}{K}\times d}$ are generated independently of $\mathbf{X}$. Each element of $\{\mathbf{Z}_{i}\}_{i=K}^{K+T-1}$ is selected uniformly i.i.d. from field $\mathbb{F}$. In addition, $N$ distinct values $\{\alpha_i\}^{N-1}_{i=0}$ over field $\mathbb{F}$ need to be selected while satisfying $\{\alpha_i\}^{N-1}_{i=0}\cup\{\beta_i\}_{i=0}^{K+T-1}=\varnothing$. Thus, the master executes $\tilde{\mathbf{X}}_i=u(\alpha_i)$ to obtain the encoded data $\{\tilde{\mathbf{X}}_i\}_{i\in\mathcal{N}}$.
Furthermore, it is noteworthy that $u(\beta_i)=\mathbf{X}_i$ for $i\in\mathcal{K}\triangleq\{0,1,\ldots,K-1\}$.

To prevent data from being eavesdropped during transmission, the encoded data $\{\tilde{\mathbf{X}}_i\}_{i=0}^{N-1}$ can be encrypted into ciphertext $\{\mathbf{C}_i\}_{i=0}^{N-1}$ by
the master utilizing the MEA-ECC algorithm. Following that, the master distributes the $\mathbf{C}_i$ to worker $W_i$ for $i\in\mathcal{N}$.

\begin{algorithm}[!t]
\small
\label{algSPCDC}
\caption{SPACDC Scheme}
\KwIn{$m, d, \mathbf{X}, T, K, N$}
\KwOut{$\{\mathbf{Y}_i\}_{i=0}^{K-1}$}
[~\uppercase\expandafter{\romannumeral1}~]~\textbf{Data process:}~the master encodes $\mathbf{X}$\;
$\mathbf{X}$ is divided into $K$ block matrices by the master\;
\For{$i=0:N-1$}
{
    $\tilde{\mathbf{X}}_i=\sum_{i=0}^{K-1}\frac{(-1)^i}{(z-\beta_i)\Gamma(z)}\mathbf{X}_i+\sum_{i=K}^{K+T-1}\frac{(-1)^i}{(z-\beta_i)\Gamma(z)}\mathbf{Z}_i$\;
    The master encrypts $\tilde{\mathbf{X}}_i$ into $\mathbf{C}_i$\;
    The master distributes $\mathbf{C}_i$ to $W_i$\;
}
[~\uppercase\expandafter{\romannumeral2}~]~\textbf{Task Computing:} $W_i$ executes $\tilde{\mathbf{Y}}_i=f(\tilde{\mathbf{X}}_i)$\;
\For{$i=0:N-1$}
{
    \If{$W_i$ \rm{receives} $\mathbf{C}_i$}
    {
        $W_i$ obtains $\tilde{\mathbf{X}}_i$ by decrypting $\mathbf{C}_i$\;
        $W_i$ performs $\tilde{\mathbf{Y}}_i=f(\tilde{\mathbf{X}}_i)$\;
        $W_i$ encrypts $\tilde{\mathbf{Y}}_i$ into $\tilde{\mathbf{C}}_i$\;
        $W_i$ returns $\tilde{\mathbf{C}}_i$ back to the master\;
    }
}
[~\uppercase\expandafter{\romannumeral3}~]~\textbf{Result Recovering:} the master recovers $\mathbf{Y}_i$\;
The master decrypts $\tilde{\mathbf{C}}_i$\;
The master constructs $(\alpha_i,f(u(\alpha_i))$ for $i\in\mathcal{F}$\;
The master designs the decoding function $h(z)=\sum_{i\in\mathcal{F}}\frac{\frac{(-1)^i}{z-\alpha_i}}{\sum_{j\in\mathcal{F}}\frac{(-1)^j}{z-\alpha_j}}f(u(\alpha_i))$\;
The master executes $\mathbf{Y}_i=f(\mathbf{X}_i)\approx h(\beta_i)$ for $i\in\{0,1,\ldots,K-1\}$\;
\textbf{Return:}{~$\mathbf{Y}_i$}
\end{algorithm}

\subsubsection{\textbf{Task Computing}}
When the $\mathbf{C}_i$ is received by worker $W_i$, $\mathbf{C}_i$ is decrypted by utilizing the MEA-ECC to obtain the original data $\tilde{\mathbf{X}}_i$.
Then, worker $W_i$ executes task $\tilde{\mathbf{Y}}_i=f(\tilde{\mathbf{X}}_i)$. Before sending the computation result back to the master, the worker uses the MEA-ECC algorithm to encrypt the result $\tilde{\mathbf{Y}}_i$ into ciphertext $\tilde{\mathbf{C}}_i$.
\subsubsection{\textbf{Result Recovering}}
The master receives data $\tilde{\mathbf{C}}_i$ from worker $W_i$ and uses the MEA-ECC algorithm to decrypt $\tilde{\mathbf{C}}_i$ to obtain the original computational result $\tilde{\mathbf{Y}}_i$. After that, we design a decoding function based on received points $(\alpha_i,f(u(\alpha_i))$, is given by
\begin{equation}
\begin{split}
\label{eqDecFunc}
h(z)=\sum_{i\in\mathcal{F}}\frac{\frac{(-1)^i}{z-\alpha_i}}{\sum_{j\in\mathcal{F}}\frac{(-1)^j}{z-\alpha_j}}f(u(\alpha_i)).
\end{split}
\end{equation}
where $i\in\mathcal{F}$ and $\mathcal{F}$ represents the set of indexes for the  workers who complete their assigned tasks.

From Eq.~\eqref{eqDecFunc}, the master can compute an approximation of $\mathbf{Y}_i=f(\mathbf{X}_i)\approx h(\beta_i)$ for $i\in\mathcal{K}$.
Moreover, the detailed steps for implementing the SPACDC scheme can be found in Algorithm~\ref{algSPCDC}.

\section{Application: The SPACDC Scheme For Deep Learning}
In this section, we introduce the SPACDC-based DL algorithm, named SPACDC-DL, which integrates the proposed SPACDC scheme into the training process of deep learning to reduce the training time while guaranteeing security and privacy of the training dataset in distributed computing scenarios.
\subsection{Problem Formulation}
\begin{figure}[!t]
\centering
\includegraphics[width=3.3in]{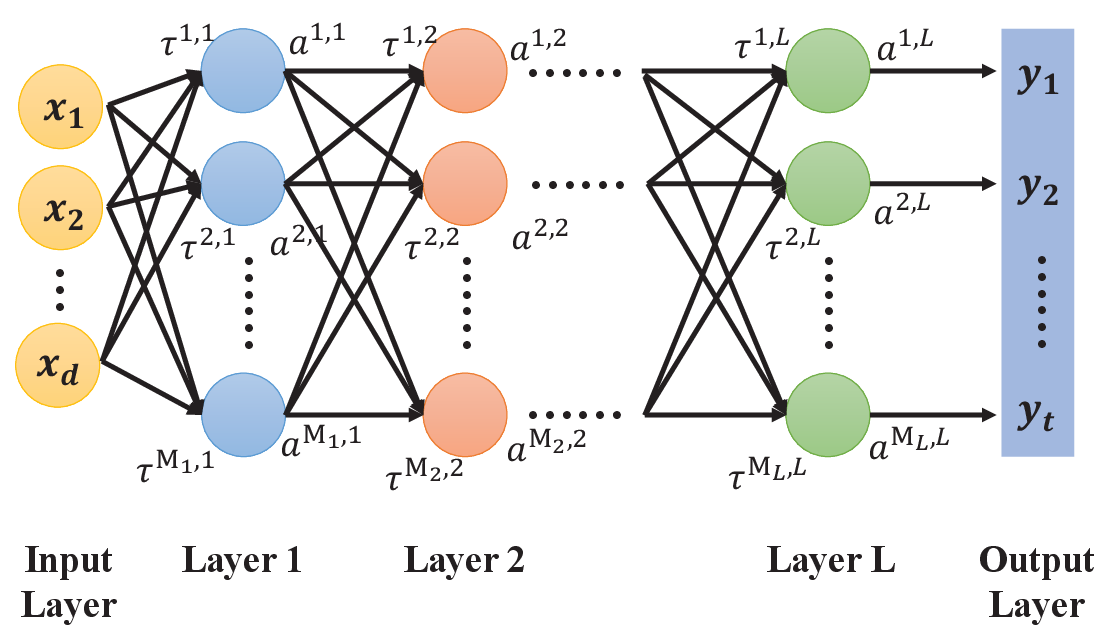}
\caption{Visualized example of a DNN with $L$ layers, each of which has $M_l$ neurons for $l=1,2,\ldots,L$.}
\label{fig:DNN}
\end{figure}

As depicted in Fig.~\ref{fig:DNN}, we focus on the issue of training a DNN model with $L$ layers, each of which has $M_l$ neurons for $l=1,2,\ldots,L$. We define $\mathcal{L}\triangleq\{1,2,\ldots,L\}$. Then, the DNN consists of $L$ weight matrices and bias vectors,
i.e., $\mathbf{\Theta}=\{\mathbf{\Theta}^1,\mathbf{\Theta}^2,\ldots,\mathbf{\Theta}^L\}$ and $\mathbf{b}=\{\mathbf{b}^1,\mathbf{b}^2,\ldots,\mathbf{b}^L\}$. Let $\mathbf{a}^l\in\mathbb{F}^{M_l}$ denote the
$l$th layer output vector, which is defined by
\begin{equation}
\begin{split}
\mathbf{a}^l=\sigma(\bm{\tau}^l)=\sigma(\mathbf{\Theta}^l\mathbf{a}^{l-1}+\mathbf{b}^l),
\end{split}
\end{equation}
where $\sigma(\cdot)$ is an activation function, and $\mathbf{b}^l$ is the bias vector of the $l$th layer.

We assume that the training dataset, denoted by $\mathcal{D}$, consists of $m$ data points. We have $\mathcal{D}=\{(\mathbf{x}_i,\mathbf{y}_i)\}_{i=1}^m$, where $\mathbf{x}_i\in\mathbb{F}^{d}$ denotes the $i$-th input feature vector and $\mathbf{y}_i$
denotes the label. The goal is to obtain the optimal weight matrices $\{\mathbf{\Theta}_i\in\mathbb{F}^{M_l\times M_{l-1}}\}_{i=1}^L$ by minimizing the loss function:
\begin{equation}
\begin{split}
J(\mathbf{\Theta};\mathbf{b};\mathbf{x}_i,\mathbf{y}_i)=\frac{1}{2|\mathcal{D}|}\sum_{(\mathbf{x}_i,\mathbf{y}_i)\in\mathcal{D}}\|\mathbf{a}^{i,L}-\mathbf{y}_i\|.
\label{lossFunc}
\end{split}
\end{equation}
Generally, we solve the above optimization problem by the gradient descent algorithm. The $l$th layer weight matrix $\mathbf{\Theta}^l$ and bias vector $\mathbf{b}^l$ are updated iteratively as
follows:
\begin{equation}
\begin{split}
\mathbf{\Theta}^l&=\mathbf{\Theta}^l-\eta\sum_{i=1}^{m}\bm{\delta}^{i,l}(\mathbf{a}^{i,l-1})^T,\\
\mathbf{b}^l&=\mathbf{b}^l-\eta\sum_{i=1}^{m}\bm{\delta}^{i,l},
\label{iterationEq}
\end{split}
\end{equation}
where $\eta$ represents the learning rate, $\odot$ denotes Hadamard product, and $\bm{\delta}^{i,l}$ is defined by
\begin{equation}
\begin{split}
\bm{\delta}^{i,l}=(\mathbf{\Theta}^l)^T\bm{\delta}^{i,l+1}\odot\sigma'(\bm{\tau}^{i,l}).
\label{deltaEq}
\end{split}
\end{equation}

From Eq.~\eqref{iterationEq} and Eq.~\eqref{deltaEq}, we find that the update process of the model parameters involves numerous matrix-vector multiplication operations.
Clearly, the explosive growth of datasets and large-scale model parameters significantly prolong the training time, which becomes a crucial challenge in the training of DNNs. In the next
section, we design a novel SPACDC-based DL algorithm to cope with these issues.

\subsection{SPACDC Scheme for Deep Learning}
Now, we give the specific description of the proposed SPACDC-DL algorithm based on a distributed computing framework. The task of the master consists of two parts: (i) computing Eq.~\eqref{deltaEq} and (ii) substituting the computed result of Eq.~\eqref{deltaEq} into equation Eq.~\eqref{iterationEq}.

According to Eq.~\eqref{deltaEq}, we define the function to be computed by the master as
\begin{equation}
f_{\delta}(\mathbf{\Theta}^l)=(\mathbf{\Theta}^l)^T\bm{\delta}^{i,l+1}\odot\sigma'(\bm{\tau}^{i,l}).
\label{deltaEq2}
\end{equation}

In each iteration, the master evenly partitions the parameter matrix $\mathbf{\Psi}\in\mathbb{F}^{M_{l-1}\times M_{l}}$ into $K$ block matrices by row, i.e.,
\begin{equation}
\label{eqDivWeightMtx}
\mathbf{\Theta}^l=
\begin{bmatrix}
\mathbf{\Theta}^l_0 \\
\mathbf{\Theta}^l_1 \\
\vdots\\
\mathbf{\Theta}^l_{K-1}
\end{bmatrix},
\end{equation}
where $\mathbf{\Theta}^l_i\in\mathbb{F}^{\frac{M_{l-1}}{K}\times M_{l}}$ for $i\in\mathcal{K}$ and $l\in\mathcal{L}$. If $M_{l-1}$ is not divisible by $K$, the final block may be zero-padded. Hence, the master's task is transformed into computing an approximation of $\mathbf{\Lambda}_i\approx f_{\delta}(\mathbf{\Theta}^l_i)$ for $i\in\mathcal{K}$. After that, the submatrices blocks $\{\mathbf{\Theta}^l_i\}^{K-1}_{i=0}$ are encoded by the master utilizing the designed encoding function:
\begin{equation}
\begin{split}
\label{eqEndFunc2}
v(z)=\sum_{i=0}^{K-1}\frac{(-1)^i}{(z-\xi_i)\Upsilon(z)}\mathbf{\Theta}^l_i+\sum_{i=K}^{K+T-1}\frac{(-1)^i}{(z-\xi_i)\Upsilon(z)}\mathbf{\Omega}^l_i,
\end{split}
\end{equation}
where $\Upsilon(z)=\sum_{j=0}^{K+T-1}\frac{(-1)^j}{z-\xi_j}$, $\{\xi_1,\xi_2,\ldots,\xi_{K+T-1}\}$ are distinct values selected from $\mathbb{F}$ by the master. Random matrices $\{\mathbf{\Omega}^l_{i}\}_{i=K}^{K+T-1}\in\mathbb{F}^{\frac{M_{l-1}}{K}\times M_{l}}$ are generated independently of $\mathbf{\Theta}$ by the master. Each element of $\{\mathbf{\Omega}^l_{i}\}_{i=K}^{K+T-1}$ is selected uniformly i.i.d. from field $\mathbb{F}$. Following that, $N$ distinct values $\{\epsilon_i\}^{N-1}_{i=0}$ over field $\mathbb{F}$ need to be selected while satisfying $\{\epsilon_i\}^{N-1}_{i=0}\cup\{\xi_i\}_{i=0}^{K+T-1}=\varnothing$. Thus, the master executes $\tilde{\mathbf{\Theta}}^l_i=v(\epsilon_i)$ for $i\in\mathcal{N}$ to complete the data encoding. In addition, it is worth mentioning that $v(\xi_i)=\mathbf{\Theta}^l_i$ for $i\in\mathcal{K}$. To prevent data from being eavesdropped during transmission, the master uses the proposed MEA-ECC to encrypt the encoded data $\{\tilde{\mathbf{\Theta}}^l_i\}_{i=0}^{N-1}$ into ciphertext $\{\mathbf{\Xi}_i\}_{i=0}^{N-1}$. After that, the encrypted data $\mathbf{\Xi}_i$ is sent to worker $W_i$ for $i\in\mathcal{N}$.

Having received the encrypted data $\mathbf{\Xi}_i$, worker $W_i$ decrypts $\mathbf{\Xi}_i$ to obtain the encoded data $\tilde{\mathbf{\Theta}}^l_i$ using its own private key. Next,
worker $W_i$ executes $\tilde{\mathbf{\Lambda}}_i=f_{\delta}(\tilde{\mathbf{\Theta}}^l_i)$. Before sending the computational result back to the master, the worker uses the MEA-ECC algorithm to encrypt the result $\tilde{\mathbf{\Lambda}}_i$ into ciphertext $\tilde{\mathbf{\Xi}}_i$.

The master receives data $\tilde{\mathbf{\Xi}}_i$ from worker $W_i$ and uses its private key to decrypt $\tilde{\mathbf{\Xi}}_i$ to obtain the original computational result $\tilde{\mathbf{\Lambda}}_i$. Let $\mathcal{G}$ represent the set of indexes for the workers who send the task results back to the master. Then, we design a decoding
function $\aleph(z)$ to approximately interpolates $f_{\delta}(v(z))$ by points $\big(\epsilon_i,f_{\delta}(v(\epsilon_i))\big)$ for $i\in\mathcal{G}$, is given by
\begin{equation}
\begin{split}
\label{eqDecFunc2}
\aleph(z)=\sum_{i\in\mathcal{G}}\frac{\frac{(-1)^i}{z-\epsilon_i}}{\sum_{j\in\mathcal{G}}\frac{(-1)^j}{z-\epsilon_j}}f_{\delta}(v(\epsilon_i)).
\end{split}
\end{equation}
Consequently, the master can obtain an approximation of $\mathbf{\Lambda}_i=f_{\delta}(\mathbf{\Theta}^l_i)\approx\aleph(\xi_i)$ for $i\in\mathcal{K}$.

The detailed steps for implementing the SPACDC-based DL scheme can be found in Algorithm~\ref{algSPCDCForDNN}.

\begin{algorithm}[!t]
\small
\label{algSPCDCForDNN}
\caption{SPACDC-based DL Algorithm}
\KwIn{$\mathcal{D}, K, N, T, L, m, d$}
\textbf{Initialization:}\\
Initialize weight matrix $\mathbf{\Theta}$ and bias vector $\mathbf{b}$\;
\For{\rm{epoch} $t=0:T_{\rm{max}}$}
{
    \For{$i=1:m$}
    {
        Set $\mathbf{a}^1$ as $\mathbf{x}_1$\;
        \For{$l=2:L$}
        {
            $\mathbf{a}^{i,l}=\sigma(\bm{\tau}^{i,l})=\sigma(\mathbf{\Theta}^l\mathbf{a}^{i,l-1}+\mathbf{b}^l)$\;
        }
        Start to minimizing the loss function~\eqref{lossFunc}\;
        \For{$l=L-1:2$}
        {
            Compute Eq.~\eqref{deltaEq2} by our SPACDC Algorithm~\ref{algSPCDC}\;
        }
        Update weight matrix and bias vector\;
        \For{$l=L-1:1$}
        {
            $\mathbf{\Theta}^l=\mathbf{\Theta}^l-\eta\sum_{i=1}^{m}\bm{\delta}^{i,l}(\mathbf{a}^{i,l-1})^T$\;
            $\mathbf{b}^l=\mathbf{b}^l-\eta\sum_{i=1}^{m}\bm{\delta}^{i,l}$\;
        }
    }
}
\KwOut{$\mathbf{\Theta}, \mathbf{b}$}
\end{algorithm}

\section{Convergence Analysis and Experiments}
In this section, we discuss the convergence of the SPACDC-DL algorithm. Then, extensive experiments are implemented to verify the superiority of the SPACDC-DL algorithm.

\subsection{Convergence Analysis}
Firstly, we give the following assumptions:

\textit{\textbf{Assumption}~1}:~For constant $\mathcal{L}>0$, the gradient function $\triangledown J(\mathbf{\Theta})$ is Lipschitz continuous, i.e.,
\begin{equation}
\begin{split}
\parallel J(\mathbf{\Theta})-J(\mathbf{\Theta}')\parallel\leq\mathcal{L}\parallel\mathbf{\Theta}-\mathbf{\Theta}'\parallel,~\textrm{for~all}~\mathbf{\Theta},\mathbf{\Theta}'.
\label{eqLips}
\end{split}
\end{equation}

\textit{\textbf{Assumption}~2}:~$\boldsymbol{\psi}^{(t)}$ is an unbiased estimate of the true gradient of the loss function $J(\mathbf{\Theta})$, i.e.,
\begin{equation}
\begin{split}
\mathbb{E}[\boldsymbol{\psi}^{(t)}]=\triangledown J(\mathbf{\Theta}^{(t)}),~\textrm{for~all}~t,
\label{eqassump2}
\end{split}
\end{equation}
where $\boldsymbol{\psi}^{(t)}$ satisfies $\mathbf{\Theta}^{(t+1)}=\mathbf{\Theta}^{(t)}-\eta\boldsymbol{\psi}^{(t)}$.

\textit{\textbf{Assumption}~3}:~There exists a scalar $\varphi>0$ such that
\begin{equation}
\begin{split}
\mathbb{E}[\parallel\boldsymbol{\psi}^{(t)}-\mathbb{E}[\boldsymbol{\psi}^{(t)}]\parallel^2]\leq\varphi^2,~\textrm{for~all}~t,
\end{split}
\end{equation}
where $\parallel\cdot\parallel$ denotes $l_2$-norm.

Based on above assumptions, we discuss the convergence of the DNN training using the proposed SPACDC-DL algorithm.

\begin{thm}
\label{thmo0}
Given a CDC system with a master and $N$ workers, the SPACDC-DL algorithm is applied to train a DNN model while satisfying \textit{Assumptions}~$1,2$ and $3$. Then, the SPACDC-DL guarantees
\begin{equation}
\mathbb{E}[J(\frac{1}{\Gamma}\sum_{t=0}^{\Gamma}\mathbf{\Theta}^{(t)})]-J(\mathbf{\Theta}^{(\ast)})\leq\frac{1}{2\eta\Gamma}\parallel\mathbf{\Theta}^{(0)}-\mathbf{\Theta}^{\ast}\parallel^2,
\end{equation}
where $\varphi$ is given in \textit{Assumption}~3, $\Gamma$ represents the number of iterations, and $\mathbf{\Theta}^{(\ast)}$ is one of the optimal parameter.
\end{thm}

\begin{IEEEproof}
From Eqs.~\eqref{lossFunc}, \eqref{eqLips} and using Taylor's formula, we have
\begin{equation}
\begin{split}
J(\mathbf{\Theta}^{(t+1)})\leq& J(\mathbf{\Theta}^{(t)})+\langle\triangledown J(\mathbf{\Theta}^{(t)}),\mathbf{\Theta}^{(t+1)}-\mathbf{\Theta}^{(t)}\rangle\\
&+\frac{\mathcal{L}}{2}\parallel\mathbf{\Theta}^{(t+1)}-\mathbf{\Theta}^{(t)}\parallel^2\\
=&J(\mathbf{\Theta}^{(t)})-\eta\langle\triangledown J(\mathbf{\Theta}^{(t)}),\boldsymbol{\psi}^{(t)}\rangle+\frac{\mathcal{L}\eta^{2}}{2}\parallel\boldsymbol{\psi}^{(t)}\parallel^2,
\label{eq32}
\end{split}
\end{equation}
where $\langle\cdot,\cdot\rangle$ denotes the inner product, i.e.,
\begin{equation}
\langle a_1,a_2\rangle=\frac{1}{2}(\parallel a_1 \parallel^2+\parallel a_2 \parallel^2-\parallel a_1-a_2 \parallel^2).
\end{equation}
Taking the expectation for both sides of Eq.~\eqref{eq32}, we obtain
\begin{equation}
\begin{split}
&\mathbb{E}[J(\mathbf{\Theta}^{(t+1)})]\\
&\leq \mathbb{E}\bigg[J(\mathbf{\Theta}^{(t)})-\eta\langle\triangledown J(\mathbf{\Theta}^{(t)}),\boldsymbol{\psi}^{(t)}\rangle+\frac{\mathcal{L}\eta^{2}}{2}\parallel\boldsymbol{\psi}^{(t)}\parallel^2\bigg]\\
&=J(\mathbf{\Theta}^{(t)})-\eta\mathbb{E}[\langle\triangledown J(\mathbf{\Theta}^{(t)}),\boldsymbol{\psi}^{(t)}\rangle]+\frac{\mathcal{L}\eta^{2}}{2}\mathbb{E}[\parallel\boldsymbol{\psi}^{(t)}\parallel^2].\\
\label{eq34}
\end{split}
\end{equation}
From Eq.~\eqref{eqassump2}, we have
\begin{equation}
\mathbb{E}[\langle\triangledown J(\mathbf{\Theta}^{(t)}),\boldsymbol{\psi}^{(t)}\rangle]=\parallel\triangledown J(\mathbf{\Theta}^{(t)}\parallel^2.
\label{eq35}
\end{equation}
For $\mathbb{E}[\parallel\boldsymbol{\psi}^{(t)}\parallel^2]$, we have
\begin{equation}
\begin{split}
\mathbb{E}[\parallel\boldsymbol{\psi}^{(t)}\parallel^2]&=\mathbb{E}[\parallel\boldsymbol{\psi}^{(t)}-\triangledown J(\mathbf{\Theta}^{(t)})+\triangledown J(\mathbf{\Theta}^{(t)})\parallel^2]\\
&=\mathbb{E}[\parallel\boldsymbol{\psi}^{(t)}-\triangledown J(\mathbf{\Theta}^{(t)})\parallel^2]+\parallel\triangledown J(\mathbf{\Theta}^{(t)})\parallel^2\\
&\leq\varphi^2+\parallel\triangledown J(\mathbf{\Theta}^{(t)})\parallel^2.
\label{eq36}
\end{split}
\end{equation}
Substituting Eqs.~\eqref{eq35} and \eqref{eq36} into Eq.~\eqref{eq34}, we get
\begin{equation}
\begin{split}
&\mathbb{E}[J(\mathbf{\Theta}^{(t+1)})]\\
&\leq J(\mathbf{\Theta}^{(t)})-\eta\parallel\triangledown J(\mathbf{\Theta}^{(t)}\parallel^2+\frac{\mathcal{L}\eta^{2}}{2}(\parallel\triangledown J(\mathbf{\Theta}^{(t)})\parallel^2+\varphi^2)\\
&=J(\mathbf{\Theta}^{(t)})-\eta(1-\frac{\mathcal{L}\eta}{2})\parallel\triangledown J(\mathbf{\Theta}^{(t)})\parallel^2+\frac{\mathcal{L}\eta^2\varphi^2}{2}.
\label{eq37}
\end{split}
\end{equation}
Due to $\mathcal{L}\eta\leq 1$, we obtain
\begin{equation}
\begin{split}
\mathbb{E}[J(\mathbf{\Theta}^{(t+1)})]\leq J(\mathbf{\Theta}^{(t)})-\frac{\eta}{2}\parallel\triangledown J(\mathbf{\Theta}^{(t)})\parallel^2+\frac{\eta\varphi^2}{2}.
\label{eq38}
\end{split}
\end{equation}
Considering the convexity of $J(\mathbf{\Theta})$, we have
\begin{equation}
J(\mathbf{\Theta}^{(t)})\leq J(\mathbf{\Theta}^{\ast})+\langle\triangledown J(\mathbf{\Theta}^{(t)}),\mathbf{\Theta}^{(t)}-\mathbf{\Theta}^{\ast}\rangle.
\label{eq39}
\end{equation}
Substituting Eq.~\eqref{eq39} into Eq.~\eqref{eq38}, we can obtain
\begin{equation}
\begin{split}
\mathbb{E}[J(\mathbf{\Theta}^{(t+1)})]&\leq J(\mathbf{\Theta}^{\ast})+\langle\triangledown J(\mathbf{\Theta}^{(t)}),\mathbf{\Theta}^{(t)}-\mathbf{\Theta}^{\ast}\rangle\\
&\quad-\frac{\eta}{2}\parallel\triangledown J(\mathbf{\Theta}^{(t)})\parallel^2+\frac{\eta\varphi^2}{2}.
\label{eq40}
\end{split}
\end{equation}
Taking the expectation for both sides of Eq.~\eqref{eq40} and combining it with Eq.~\eqref{eqassump2}, we get
\begin{equation}
\begin{split}
\mathbb{E}[J(\mathbf{\Theta}^{(t+1)})]&\leq J(\mathbf{\Theta}^{\ast})+\mathbb{E}[\langle\boldsymbol{\psi}^{(t)},\mathbf{\Theta}^{(t)}-\mathbf{\Theta}^{\ast}\rangle]\\
&\quad-\frac{\eta}{2}\mathbb{E}[\parallel\boldsymbol{\psi}^{(t)}\parallel^2]+\frac{\eta\varphi^2}{2}.
\label{eq41}
\end{split}
\end{equation}
Substituting Eq.~\eqref{eq36} into Eq.~\eqref{eq41}, we obtain
\begin{equation}
\begin{split}
&\mathbb{E}[J(\mathbf{\Theta}^{(t+1)})]\\
&\leq J(\mathbf{\Theta}^{\ast})+\mathbb{E}[\langle\boldsymbol{\psi}^{(t)},\mathbf{\Theta}^{(t)}-\mathbf{\Theta}^{\ast}\rangle]+\frac{\eta\varphi^2}{2}\\
&\quad-\frac{\eta}{2}(\varphi^2+\parallel\triangledown J(\mathbf{\Theta}^{(t)})\parallel^2)\\
&=J(\mathbf{\Theta}^{\ast})+\mathbb{E}[\langle\boldsymbol{\psi}^{(t)},\mathbf{\Theta}^{(t)}-\mathbf{\Theta}^{\ast}\rangle]-\frac{\eta}{2}\parallel\triangledown J(\mathbf{\Theta}^{(t)})\parallel^2\\
&=J(\mathbf{\Theta}^{\ast})+\mathbb{E}\bigg[\langle\boldsymbol{\psi}^{(t)},\mathbf{\Theta}^{(t)}-\mathbf{\Theta}^{\ast}\rangle-\frac{\eta}{2}\parallel\boldsymbol{\psi}^{(t)}\parallel^2\bigg]\\
&=J(\mathbf{\Theta}^{\ast})+\frac{1}{2\eta}(\mathbb{E}\parallel\mathbf{\Theta}^{(t)}-\mathbf{\Theta}^{\ast}\parallel^2-\mathbb{E}\parallel\mathbf{\Theta}^{(t+1)}-\mathbf{\Theta}^{\ast}\parallel^2).
\label{eq42}
\end{split}
\end{equation}
By summing Eq.~\eqref{eq42} for $t=1,2,\ldots,\Gamma-1$, we get
\begin{equation}
\begin{split}
&\sum_{t=0}^{\Gamma-1}(\mathbb{E}[J(\mathbf{\Theta}^{(t+1)})]-J(\mathbf{\Theta}^{\ast}))\\
&\leq\frac{1}{2\eta}(\mathbb{E}\parallel\mathbf{\Theta}^{(0)}-\mathbf{\Theta}^{\ast}\parallel^2-\mathbb{E}\parallel\mathbf{\Theta}^{(\Gamma)}-\mathbf{\Theta}^{\ast}\parallel^2)\\
&\leq\frac{1}{2\eta}\parallel\mathbf{\Theta}^{(0)}-\mathbf{\Theta}^{\ast}\parallel^2.
\label{eq43}
\end{split}
\end{equation}
Due to the convexity of $J(\mathbf{\Theta})$, we have
\begin{equation}
\begin{split}
\mathbb{E}[J(\frac{1}{\Gamma}\sum_{t=0}^{\Gamma})]-J(\mathbf{\Theta}^{\ast})&\leq\frac{1}{\Gamma}\sum_{t=0}^{\Gamma-1}(\mathbb{E}[J(\mathbf{\Theta}^{(t+1)})]-J(\mathbf{\Theta}^{\ast}))\\
&\leq\frac{1}{2\eta\Gamma}\parallel\mathbf{\Theta}^{(0)}-\mathbf{\Theta}^{\ast}\parallel^2.
\label{eq44}
\end{split}
\end{equation}
This completes the proof.

\begin{table*}
\centering
\scriptsize
\caption{Comparison of complexity with other schemes}
\begin{center}
\begin{tabular}{|m{1.5cm}<{\centering}|m{1.2cm}<{\centering}|m{3.1cm}<{\centering}|m{1.4cm}<{\centering}|m{1.8cm}<{\centering}|m{1.6cm}<{\centering}|m{1.4cm}<{\centering}|m{1.4cm}<{\centering}|}
\hline
~& ~ & ~ &\multicolumn{2}{c|}{Communication Complexity} & ~ & ~ & ~\\
\cline{4-5}
\multirow{-2}{*}{Coded Scheme} & \multirow{-2}{*}{\makecell*[c]{Encoding\\ Complexity}} & \multirow{-2}{*}{Decoding Complexity} & Master to all workers & Workers to master & \multirow{-2}{*}{\makecell*[c]{Computational\\ Complexity}}& \multirow{-2}{*}{\makecell*[c]{Protect Data\\ Security}}& \multirow{-2}{*}{\makecell*[c]{Protect Data\\ Privacy}} \\
\hline
Polynomial Codes~\cite{Yu2017Polynomial} &$\mathcal{O}(mdN)$ &$\mathcal{O}(m^2\log^2K^2\log\log K^2)$ &$\mathcal{O}(mdN/K)$ &$\mathcal{O}(m^2)$ &$\mathcal{O}(dm^2/K^2)$ &No &No\\
\hline
MatDot Codes~\cite{dutta2020optimal} &$\mathcal{O}(mdN)$ &$\mathcal{O}(Km^2\log^2K\log\log K)$ &$\mathcal{O}(mdN/K)$ &$\mathcal{O}(Km^2)$ &$\mathcal{O}(dm^2/K)$ &No &No\\
\hline
SecPoly Codes~\cite{yang2019secure}  &$\mathcal{O}(mdN)$ &$\mathcal{O}(m^2\log^2K^2\log\log K^2)$ &$\mathcal{O}(mdN/K)$&$\mathcal{O}(m^2)$&$\mathcal{O}(dm^2/K^2)$ &No & Yes\\
\hline
BACC scheme~\cite{Jahani2023Berrut} &$\mathcal{O}(mdN)$&$\mathcal{O}(|\mathcal{F}|)$&$\mathcal{O}(mdN/K)$&$\mathcal{O}(m^2|\mathcal{F}|/K^2)$ &$\mathcal{O}(dm^2/K^2)$&No &No\\
\hline
LCC scheme~\cite{yu2019lagrange} &$\mathcal{O}(mdN)$&$\mathcal{O}(m^2\log^2K\log\log K)$&$\mathcal{O}(mdN/K)$&$\mathcal{O}(m^2/K)$ &$\mathcal{O}(dm^2/K^2)$& No &Yes\\
\hline
SPACDC code (Our Scheme)&$\mathcal{O}(mdN)$ &$\mathcal{O}(|\mathcal{F}|)$ &$\mathcal{O}(mdN/K)$ &$\mathcal{O}(m^2|\mathcal{F}|/K^2)$ &$\mathcal{O}(dm^2/K^2)$ &Yes & Yes\\
\hline
\end{tabular}
\label{TbCompAna}
\end{center}
\end{table*}

\end{IEEEproof}

\subsection{Experiments}
Now, we focus on conducting experiments to evaluate the effectiveness of the SPACDC-DL algorithm over a distributed computing system.

\subsubsection{Experimental Settings}
In our experiments, we implement the proposed SPACDC-DL algorithm using the MPI4PY package~\cite{dalcin2021mpi4py} in Python on a cluster of $31$ computing instances containing a
master and $N=30$ workers. In the system, we randomly select $S$ straggling workers and $T$ colluding workers over $N=30$ workers. The specific values of $S$ and $T$ will be provided later. To
simulate the straggling effects, we introduce artificial delays using the \textit{sleep()} function from the time package~\cite{Chen2023FTPipeHD}.

We focus on training a DNN for image classification on the popular MNIST dataset using the proposed SPACDC-DL algorithm. The CNN is composed of an input layer, three convolutional layers, two pooling
layers, two fully connected layers, and an output layer. To demonstrate the performance of the SPACDC-DL algorithm, we conduct experiments by selecting different number of stragglers $S$ based on the following four system settings:
\begin{itemize}
\item [$\bullet$]
\textbf{Scenario 1:}~$N=30, T=3$ and $S=0$.
\item [$\bullet$]
\textbf{Scenario 2:}~$N=30, T=3$ and $S=3$.
\item [$\bullet$]
\textbf{Scenario 3:}~$N=30, T=3$ and $S=5$.
\item [$\bullet$]
\textbf{Scenario 4:}~$N=30, T=3$ and $S=7$.
\end{itemize}

In addition, we introduce the following baseline algorithms:
\begin{itemize}
\item [$\bullet$]
\textbf{Algorithm 1:}~This algorithm completes the training tasks using conventional distributed computing scheme, named CONV-DL.
\item [$\bullet$]
\textbf{Algorithm 2:}~This algorithm completes the training tasks using MDS-based~\cite{lee2018speeding} distributed computing scheme, named MDS-DL.
\item [$\bullet$]
\textbf{Algorithm 3:}~This algorithm completes the training tasks using MATDOT-based~\cite{dutta2020optimal} distributed computing scheme, named MATDOT-DL.
\end{itemize}

\begin{figure}[!t]
\centering
\includegraphics[width=3in]{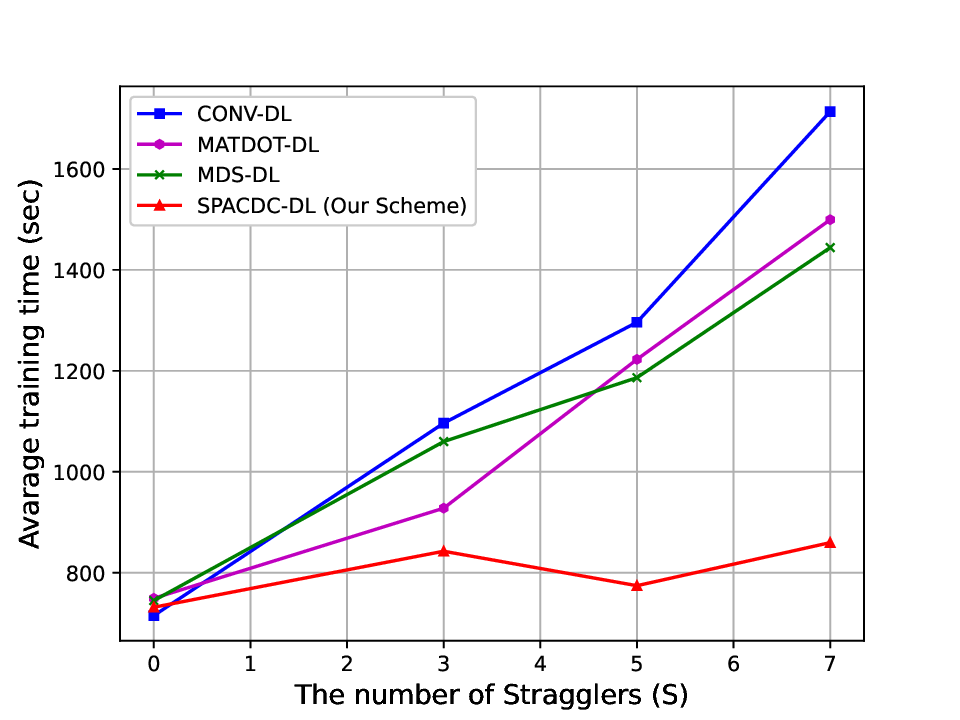}
\caption{Comparison of average training time achieved by the CONV-DL, MDS-DL, and MATDOT-DL algorithms and our proposed SPACDC-DL algorithm for training a DNN in distributed computing systems under parameters $N=30$ and $T=3$ while with $S=0, 3, 5,~\text{and}~7$ stragglers.}
\label{fig:AverageTrainingTime}
\end{figure}

\begin{figure*}
\centering
\subfigure[Stragglers $S=3$]{
\label{fg:a} 
\includegraphics[width=2.3in]{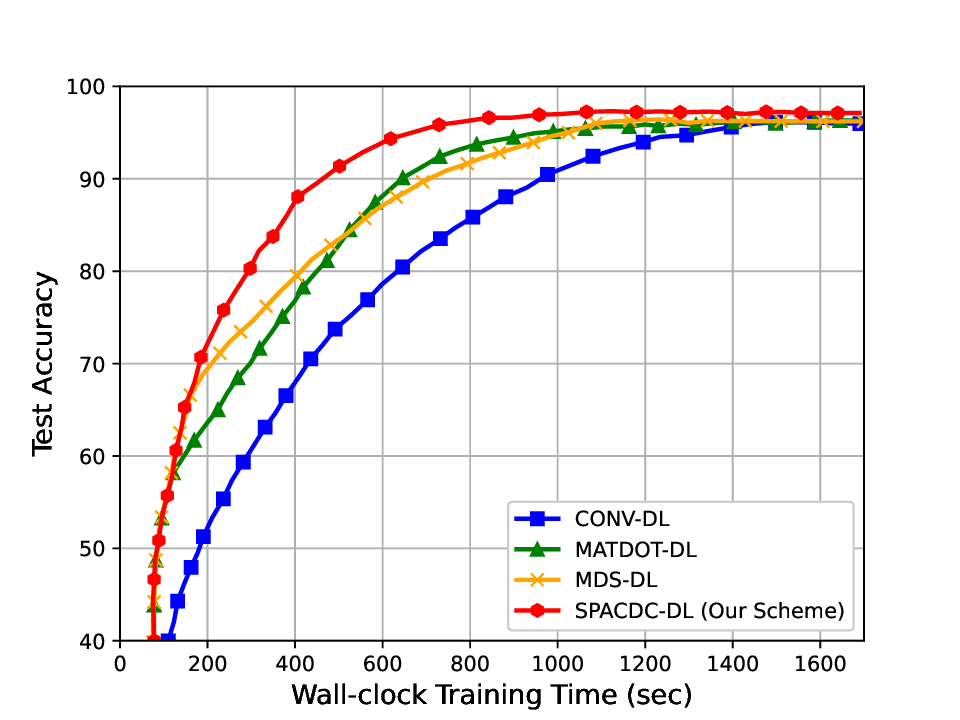}}
\subfigure[Stragglers $S=5$]{
\label{fg2:b} 
\includegraphics[width=2.3in]{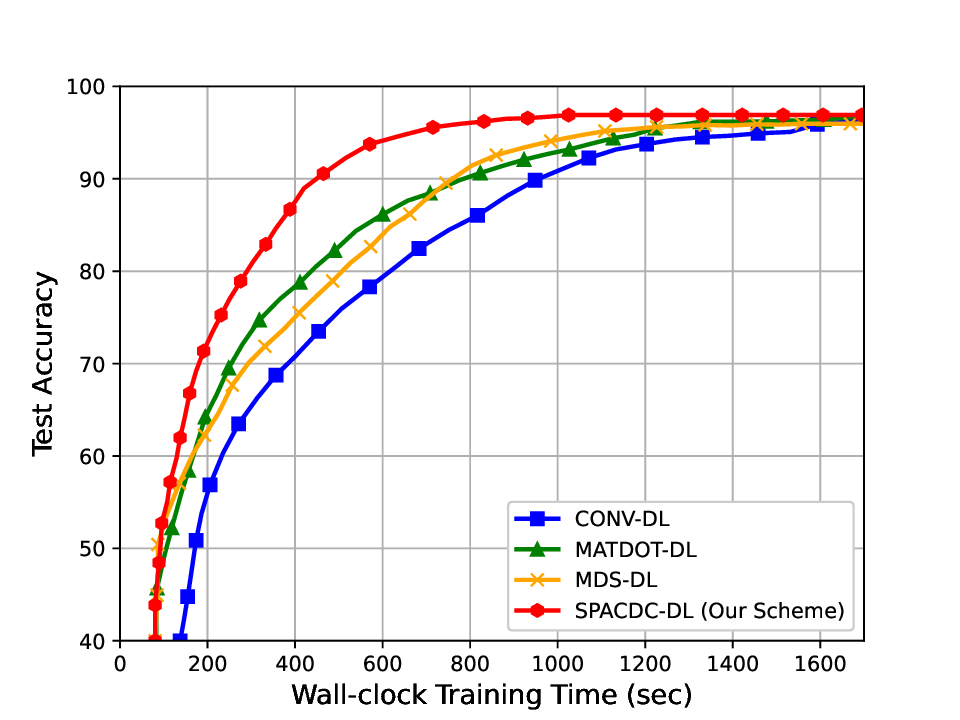}}
\subfigure[Stragglers $S=7$]{
\label{fg2:c} 
\includegraphics[width=2.3in]{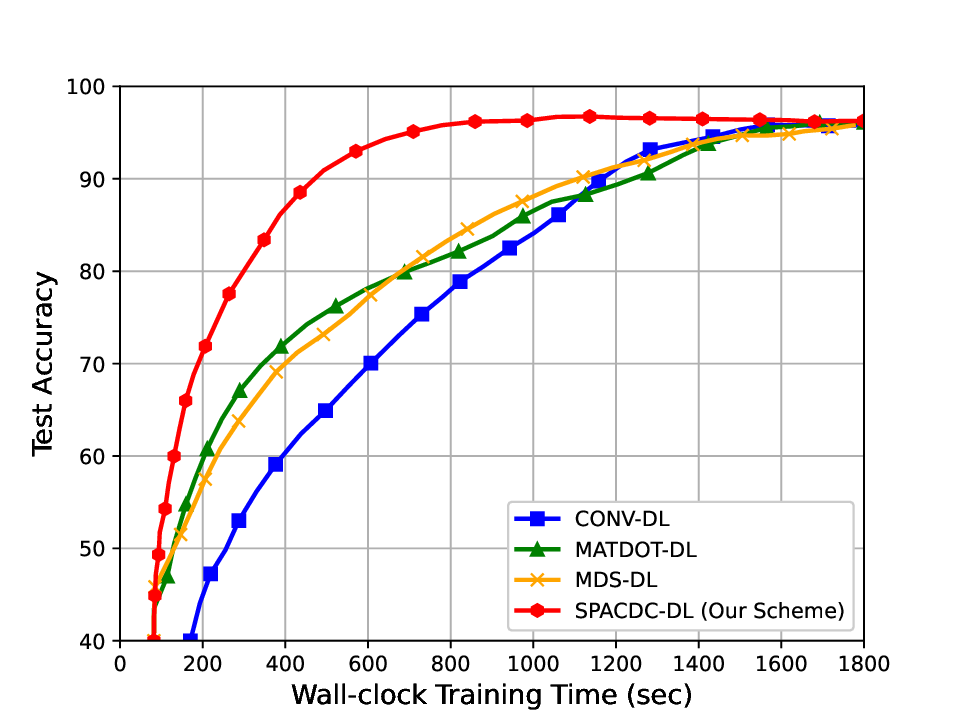}}
\caption{Comparison of test accuracy obtained by the CONV-DL, MDS-DL, and MATDOT-DL algorithms and our proposed SPACDC-DL algorithm in distributed computing systems with $N=30$, $T=3$, and $S=3,5,7$.}
\label{fig:TestAccuracy} 
\end{figure*}

\subsubsection{Experimental Results}
As illustrated in Fig.~\ref{fig:AverageTrainingTime}, we compare the average training time of the CONV-DL, MDS-DL, and MATDOT-DL algorithms and our proposed SPACDC-DL for training the DNN in
four different distributed computing scenarios, i.e., scenarios $1$, $2$, $3$, and $4$. In the experiment, we train the model $100$ times using each of the four algorithms and then take the
average of the training times. From Fig.~\ref{fig:AverageTrainingTime}, the proposed SPACDC-DL algorithm outperforms the other three algorithms with respect to average training time. When the
number of stragglers is 3, there is minimal disparity in the average training time when utilizing the four algorithms for model training. With the increasing of the number of stragglers, the average training time for the CONV-DL, MDS-DL, and MATDOT-DL algorithms increases rapidly and is much higher than that of the SPACDC-DL algorithm. The experimental results verify that our SPACDC-DL algorithm is able to provide robustness against the straggler effects in a distributed computing framework.

Figure~\ref{fig:TestAccuracy} shows the test accuracy of the CONV-DL, MDS-DL, and MATDOT-DL algorithms and our proposed SPACDC-DL for training the DNN on the MNIST dataset in a
distributed system under parameters $N=30$, $T=3$, and $S=3,5,7$. After each training epoch, we conduct $100$ tests on the test dataset and then take the average value of the test accuracy for
comparison. From Fig.~\ref{fig:TestAccuracy}, the curves show that our proposed SPACDC-DL algorithm converges faster than CONV-DL, MDS-DL, and MATDOT-DL algorithms for a given test accuracy.

Specifically, when $S=3$, the test accuracy converges to $80\%$, the proposed SPACDC-DL algorithm saves the average training time up to $54.0\%$, $26.7\%$, and $33.4\%$ compared with the
CONV-DL, MDS-DL, and MATDOT-DL algorithms, respectively. When $S=5$, the test accuracy converges to $80\%$, the proposed SPACDC-DL algorithm saves the average training time up to $52.0\%$,
$39.0\%$, and $33.6\%$ compared with the CONV-DL, MDS-DL, and MATDOT-DL algorithms, respectively. When the test accuracy converges to $90\%$, the SPACDC-DL algorithm saves the average training
time up to $54.5\%$, $43.4\%$, and $47.4\%$ compared with the CONV-DL, MDS-DL, and MATDOT-DL algorithms, respectively. When $S=7$, the test accuracy converges to $80\%$, the proposed SPACDC-DL
algorithm saves the average training time up to $65.2\%$, $54.5\%$, and $55.5\%$ compared with the CONV-DL, MDS-DL, and MATDOT-DL algorithms, respectively. When the test accuracy converges to
$90\%$, the SPACDC-DL algorithm saves the average training time up to $58.1\%$, $56.7\%$, and $62.1\%$ compared with the CONV-DL, MDS-DL, and MATDOT-DL algorithms, respectively. The MDS-DL
algorithm is slightly faster than the MATDOT-DL algorithm with respect to convergence rate, and the CONV-DL algorithm is the slowest. The convergence rate of these algorithms is mainly influenced by the recovery threshold. In addition, we observe that our proposed SPACDC-DL algorithm is superior to the CONV-DL, MDS-DL, and MATDOT-DL algorithms in terms of convergence rate.

We observe that our SPACDC-DL algorithm can offer ITP protection for the dataset during the entire training process, while other algorithms cannot protect data privacy.

\begin{figure*}
\begin{minipage}[t]{0.48\linewidth}
\centering
\includegraphics[width=3in]{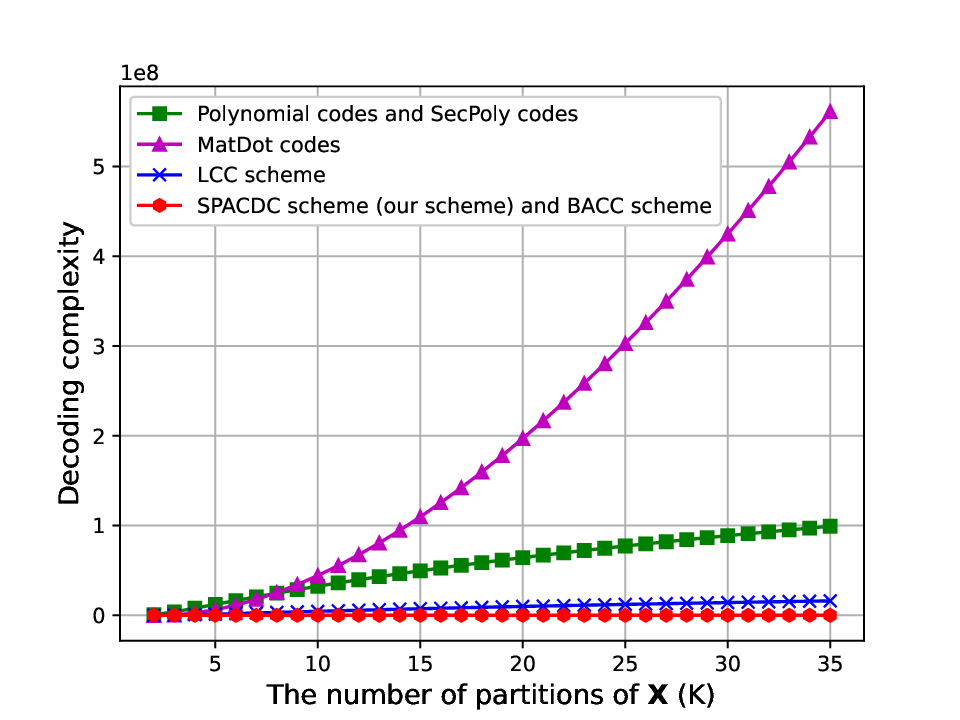}
\caption{Comparison of decoding complexity obtained by the BACC scheme, LCC scheme, Polynomial codes, SecPoly codes, MatDot codes, and the SPACDC scheme in a CDC system with parameters $m=1000$, and $K$ values range from $1$ to $36$.}
\label{fig:DecComp}
\end{minipage}
~
\begin{minipage}[t]{0.48\linewidth}
\centering
\includegraphics[width=3in]{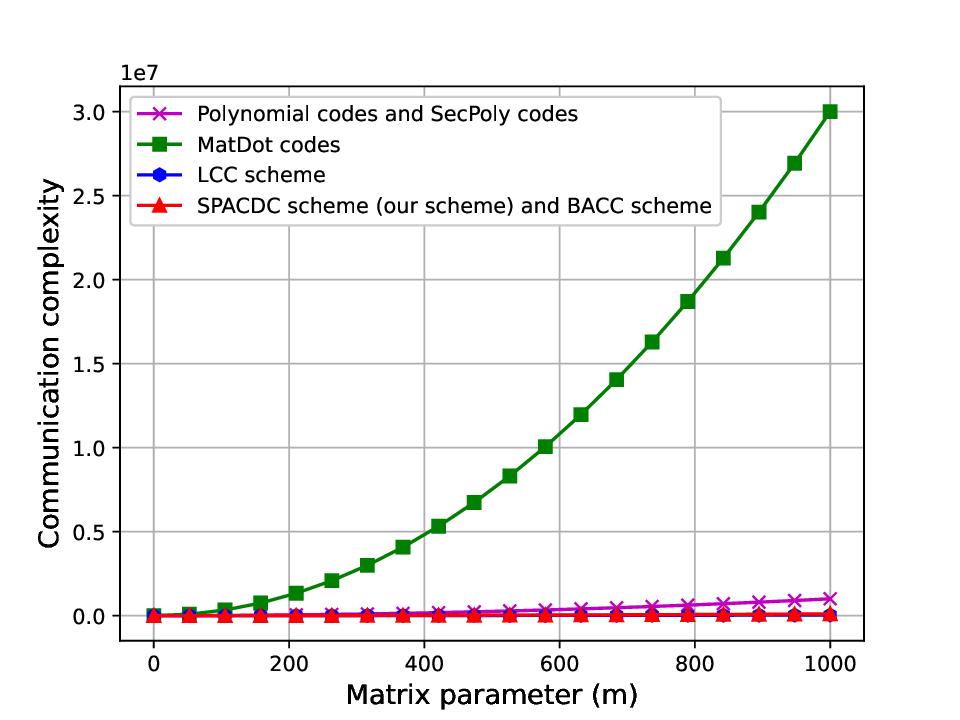}
\caption{Comparison of communication complexity of the BACC scheme, LCC scheme, Polynomial codes, SecPoly codes, MatDot codes, and the SPACDC scheme in a CDC system with parameters $|\mathcal{F}|=10$, $K=30$, and $m$ values range from $1$ to $1000$.}
\label{fig:CommComp}
\end{minipage}
\end{figure*}

\begin{figure}[!t]
\centering
\includegraphics[width=3in]{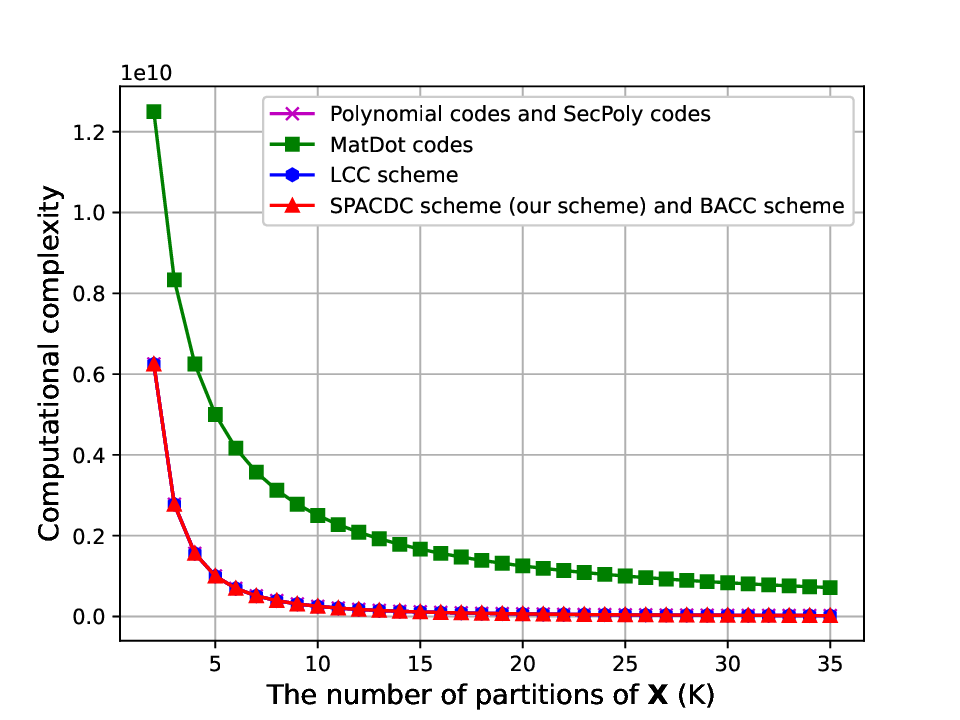}
\caption{Comparison of computational complexity of the BACC scheme, MatDot codes, Polynomial codes, LCC scheme, SecPoly codes, and our SPACDC scheme in a CDC system with parameters $d=1000$, $m=5000$, and $K$ values range from $1$ to $36$.}
\label{fig:ComputComp}
\end{figure}

\section{Result and Complexity Analyses}
In this section, we provide the theoretical proofs for our SPACDC scheme with respect to ITP. Then, we conduct a comprehensive complexity analysis of the SPACDC scheme.

\subsection{Some Theorems}
\begin{thm}
\label{thmo1}
Given a CDC system including a master and $N$ workers, worker $\{W_i\}_{i\in\mathcal{P}}$ is able to obtain no information about input matrix $\mathbf{X}$ from assigned matrix $\tilde{\mathbf{X}}_i$, i.e.,
\begin{equation}
\mathbf{I}(\tilde{\mathbf{X}}_i;\mathbf{X})=0,~\text{for}~i\in\mathcal{P}.
\end{equation}
\end{thm}
\begin{IEEEproof}
Inspired by~\cite{aliasgari2020private,qiu2024coded}, we prove Theorem~\ref{thmo1} based on information-theoretic criteria. Specifically, the mutual information for $\mathbf{X}$ and $\tilde{\mathbf{X}}_i$ is given by
\begin{subequations}
\begin{align}
&\mathbf{I}(\mathbf{X};\tilde{\mathbf{X}}_\mathcal{P}) \notag\\
&=\mathbf{H}(\tilde{\mathbf{X}}_\mathcal{P})-\mathbf{H}(\tilde{\mathbf{X}}_\mathcal{P}|\mathbf{X}) \label{eqa}\\
&=\mathbf{H}(\tilde{\mathbf{X}}_\mathcal{P}|\mathbf{X},\mathbf{Z}_{K},\mathbf{Z}_{K+1},\ldots,\mathbf{Z}_{K+T-1})\notag\\
&\quad~+\mathbf{H}(\tilde{\mathbf{X}}_\mathcal{P})-\mathbf{H}(\tilde{\mathbf{X}}_\mathcal{P}|\mathbf{X})\label{eqb}\\
&=\mathbf{H}(\tilde{\mathbf{X}}_\mathcal{P})-\mathbf{I}(\tilde{\mathbf{X}}_\mathcal{P};\mathbf{Z}_{K},\mathbf{Z}_{K+1},\ldots,\mathbf{Z}_{K+T-1}|\mathbf{X})\\
&=\mathbf{H}(\tilde{\mathbf{X}}_\mathcal{P})-\mathbf{H}(\mathbf{Z}_{K},\mathbf{Z}_{K+1},\ldots,\mathbf{Z}_{K+T-1}|\mathbf{X})\notag\\
&\quad~+\mathbf{H}(\mathbf{Z}_{K},\mathbf{Z}_{K+1},\ldots,\mathbf{Z}_{K+T-1}|\mathbf{X},\tilde{\mathbf{X}}_\mathcal{P})\\
&=\mathbf{H}(\tilde{\mathbf{X}}_\mathcal{P})-\mathbf{H}(\mathbf{Z}_{K},\mathbf{Z}_{K+1},\ldots,\mathbf{Z}_{K+T-1})\label{eqe}\\
&\leq\mathbf{H}(\tilde{\mathbf{X}}_\mathcal{P})-\sum_{i=K}^{K+T-1}\mathbf{H}(\mathbf{Z}_i)\label{eqf}\\
&=\mathbf{H}(\tilde{\mathbf{X}}_\mathcal{P})-T\frac{md}{K}\log|\mathbb{F}|\label{eqgg}\\
&\leq\sum_{i=1}^{T}\mathbf{H}(\tilde{\mathbf{X}}_i)-T\frac{md}{K}\log|\mathbb{F}|\label{eqhh}\\
&=T\frac{md}{K}\log|\mathbb{F}|-T\frac{md}{K}\log|\mathbb{F}|\label{eqii}\\
&=0,\notag
\end{align}
\end{subequations}
where \eqref{eqb} is due to the fact that $\tilde{\mathbf{X}}_\mathcal{P}$ is a deterministic function of $\mathbf{X}$ and $\{\mathbf{Z}_i\}_{i=K}^{K+T-1}$; \eqref{eqe} follows the fact that $\{\mathbf{Z}_i\}_{i=K}^{K+T-1}$ are randomly generated of $\mathbf{X}$ by the master; \eqref{eqf} and \eqref{eqhh} due to the upper bounding of the joint entropy;
\eqref{eqgg} is due to the fact that all elements of $\{\mathbf{Z}_i\}_{i=K}^{K+T-1}$ are uniformly selected from i.i.d. random variables over $\mathbb{F}$; \eqref{eqii} is derived similar to \eqref{eqgg}. This completes the proof.
\end{IEEEproof}

\begin{thm}
\label{thmo2}
Given a CDC system including a master and $N$ workers, worker $\{W_i\}_{i\in\mathcal{P}}$ is able to obtain no information about the final result $\mathbf{Y}$ from calculation sub-result $\tilde{\mathbf{Y}}_i$, i.e.,
\begin{equation}
\mathbf{I}(\tilde{\mathbf{Y}}_i;\mathbf{Y})=0,~\text{for}~i\in\mathcal{P}.
\end{equation}
\end{thm}

\begin{IEEEproof}
Similar to Theorem~\ref{thmo1}, the mutual information for $\mathbf{Y}$ and $\tilde{\mathbf{Y}}_i$ is given by
\begin{subequations}
\begin{align}
&\mathbf{I}(\mathbf{Y};\tilde{\mathbf{Y}}_i)\\
&=\mathbf{I}\big(f(\mathbf{X});f(\tilde{\mathbf{X}}_i)\big)\notag\\
&=\mathbf{H}\big(f(\mathbf{X})\big)-\mathbf{H}\big(f(\mathbf{X})|f(\tilde{\mathbf{X}}_i)\big)\label{eq22a}\\
&\leq\mathbf{H}\big(f(\mathbf{X})\big)-\mathbf{H}\big(f(\mathbf{X})|f(\tilde{\mathbf{X}}_i),\tilde{\mathbf{X}}_i\big)\label{eq22b}\\
&=\mathbf{H}\big(f(\mathbf{X})\big)-\mathbf{H}\big(f(\mathbf{X})|\tilde{\mathbf{X}}_i\big)\\
&=\mathbf{H}\big(f(\mathbf{X})\big)-\mathbf{I}\big(f(\mathbf{X});\mathbf{X}|\tilde{\mathbf{X}}_i\big)-\mathbf{H}\big(f(\mathbf{X})|\tilde{\mathbf{X}}_i,\mathbf{X}\big)\\
&=\mathbf{H}\big(f(\mathbf{X})\big)-\mathbf{I}(f(\mathbf{X});\mathbf{X}|\tilde{\mathbf{X}}_i)\label{eq22e}\\
&=\mathbf{H}\big(f(\mathbf{X})\big)-\mathbf{H}(\mathbf{X}|\tilde{\mathbf{X}}_i)+\mathbf{H}\big(\mathbf{X}|\tilde{\mathbf{X}}_i,f(\mathbf{X})\big)\\
&=\mathbf{H}\big(f(\mathbf{X})\big)-\mathbf{H}(\mathbf{X})+\mathbf{H}\big(\mathbf{X}|f(\mathbf{X})\big)\label{eq22g}\\
&=\mathbf{H}\big(f(\mathbf{X})\big)-\mathbf{I}\big(f(\mathbf{X});\mathbf{X}\big)\\
&=\mathbf{H}\big(f(\mathbf{X})|\mathbf{X}\big)\\
&=0,\label{eq22j}
\end{align}
\end{subequations}
where \eqref{eq22b} is based on the conditioning reduces entropy; \eqref{eq22e} and \eqref{eq22j} is due to the fact that $f(\mathbf{X})$ is a deterministic function of
$\mathbf{X}$; \eqref{eq22g} is derived from \eqref{eqPri} which has been proved in Theorem \ref{thmo1}. This completes the proof.
\end{IEEEproof}

\subsection{Complexity Analysis}
In this subsection, we present a detailed complexity analysis for our SPACDC scheme. We also compare our SPACDC scheme with the baseline schemes and discuss their advantages and disadvantages.
\subsubsection{Encoding Complexity}
Consider the encoding function~\eqref{eqEndFunc} for the encoding phase, we can easily find that $u(z)$ is the sum of $K+T$ matrices, and each matrix of dimension $\frac{m}{K}\times d$.
Therefore, for each worker, the encoding function $u(z)$ has a computational complexity of $\mathcal{O}(md(K+T)/K)=\mathcal{O}(md)$. The SPACDC scheme's total encoding complexity for $N$ workers is $\mathcal{O}(mdN)$.
\subsubsection{Decoding Complexity}
Consider the decoding function~\eqref{eqDecFunc} for the third phase of the SPACDC scheme, the final result $\mathbf{Y}$ is recovered by interpolating the polynomial $f(u(z))$. Following the approach in \cite{Jahani2023Berrut}, the SPACDC scheme has a decoding complexity of $\mathcal{O}(|\mathcal{F}|)$.
\subsubsection{Communication Complexity}
The SPACDC scheme's communication complexity consists of two key components: $(\textit{i})$ from master to worker and $(\textit{ii})$ from worker to master. Firstly, the master sends $\mathcal{O}(md/K)$ symbols to each worker. Therefore, the overall number of symbols sent by the master to $N$ workers totals $\mathcal{O}(mdN/K)$. Secondly, the master receives $\mathcal{O}(m^2/K^2)$ symbols from each worker. Therefore, the overall number of symbols from $|\mathcal{F}|$ workers is $\mathcal{O}(m^2|\mathcal{F}|/K^2)$.
\subsubsection{Each Worker's Computational Complexity}
Consider the computing task $\tilde{\mathbf{Y}}_i=f(\tilde{\mathbf{X}}_i)$, e.g., $\tilde{\mathbf{Y}}_i=\tilde{\mathbf{X}}_i\tilde{\mathbf{X}}^T_i$, where the dimension of $\mathbf{X}_i$ is $\frac{m}{K}\times d$. Therefore, the computational complexity for each worker is $\mathcal{O}(dm^2/K^2)$.

As illustrated in Table~\ref{TbCompAna}, We summarized the complexity analysis for our SPACDC scheme while comparing it to the baseline schemes (BACC scheme~\cite{Jahani2023Berrut}, LCC
scheme~\cite{yu2019lagrange}, polynomial codes~\cite{Yu2017Polynomial}, SecPoly codes~\cite{yang2019secure}, and MatDot codes~\cite{dutta2020optimal}) with respect to encoding, decoding,
communication, and computation. To guarantee fairness in comparisons, all encoding schemes are evaluated using the same parameter settings. From the table, it can be seen that our SPACDC scheme has similar coding complexity to the baseline schemes. Furthermore, the SPACDC scheme has the same communication complexity as the baseline schemes when it comes to communication between the master and workers. It is mainly determined by their matrix partitioning methods.

As shown in Fig.~\ref{fig:DecComp}, we provide the decoding complexity comparison of various coding schemes, including the BACC scheme, LCC scheme, Polynomial codes, SecPoly codes, MatDot
codes, and the SPACDC scheme, with parameters $m=1000$, and $K$ values range from $1$ to $36$. It is clear that our SPACDC and BACC coding schemes have the lowest decoding
complexities compared to other schemes, whereas the decoding complexity of the MatDot codes higher than all other coding schemes. The decoding complexities of the BACC, LCC and the SPACDC
scheme are lower than those of the Polynomial and SecPoly codes. Similar to the BACC scheme, our SPACDC scheme uses Berrut's rational interpolant to encode data and then greatly lowers the decoding complexity by designing a low-degree decoding function. Moreover, the decoding complexity of our SPACDC scheme is a bit lower than that of the popular LCC. Compared to our SPACDC scheme, the advantage of the LCC scheme in decoding complexity will gradually disappear with the ever-increasing degree of its encoding functions.

Figure~\ref{fig:CommComp} illustrates a comparison of the communication complexity for the BACC scheme, MatDot codes, Polynomial codes, LCC scheme, SecPoly codes, and our proposed scheme with
parameters $|\mathcal{F}|=10$ and $K=30$, $m$ values range from $1$ to $1000$. We can observe that the SPACDC and BACC schemes exhibit the lowest communication complexities. On the
contrary, MatDot codes show the highest communication complexity, from workers to the master of all the schemes. The MatDot codes' application is limited primarily by their extremely-high communication complexity. More precisely, the communication complexity is greatly impacted by the dimensions of the matrices returned by the workers.

As shown in Fig.~\ref{fig:ComputComp}, we compare the computational complexity of our SPACDC scheme and the above-mentioned schemes. For this comparison, we set system parameters as $d=1000$, $m=5000$, and $K$ values range from $1$ to $36$. The results show that, apart from MatDot codes, the SPACDC
scheme has similar computational complexity to the above-mentioned schemes. It is noteworthy that the computational complexity of the MatDot codes is higher than all other coding schemes. This
is primarily due to the matrix dimensions assigned to each worker node being too large.

As mentioned above, the proposed SPACDC scheme effectively reduces complexity in comparison to the baseline coding schemes. While the complexity of the SPACDC scheme is comparable to that of the BACC scheme, the BACC scheme cannot guarantee security and privacy of data. Hence, the proposed SPACDC scheme is superior to all other coding schemes.

\section{Conclusions}
In this work, we designed a novel coding scheme, called SPACDC scheme, which is able to approximately computing a function while guaranteeing security and privacy of the data in a distributed system based on coding theory and ECC. To prevent data from being eavesdropped during transmission, we proposed a new encryption algorithm, i.e., the MEA-ECC. In particular, we designed a secure and private distributed deep learning algorithm based on the SPACDC scheme, i.e., SPACDC-DL algorithm, to speed up the training process while providing data privacy protection. Then, the ITP protection of input data was theoretically proven. Furthermore, a detailed performance analysis was completed to validate the efficiency of our SPACDC scheme. Finally, our experiments demonstrated the superiority of our SPACDC-DL algorithm in terms of average training time and privacy protection.

\bibliographystyle{IEEEtran}
\bibliography{cite}

\end{document}